\documentclass[11pt,reqno]{amsart}
\usepackage{amsfonts,amsmath,amssymb}
\newtheorem{thm}{Theorem}[section]
\newtheorem{proposition}[thm]{Proposition}
\newtheorem{corollary}[thm]{Corollary}
\newtheorem{lemma}[thm]{Lemma}

\usepackage{graphicx}
\usepackage{epstopdf}

\title[Deformed Calogero-Moser operator of the type $BC(m,n)$ and super Jacobi polynomials ]{$BC_{\infty}$ Calogero-Moser operator and  super Jacobi polynomials }
\author{ A.N. Sergeev}\address{Department of Mathematical Sciences,
Loughborough University, Loughborough LE11 3TU, UK and Steklov Institute of Mathematics,
Fontanka 27, St. Petersburg, 191023, Russia}
\email{A.N.Sergeev@lboro.ac.uk}

\author{A.P. Veselov}
\address{Department of Mathematical Sciences,
Loughborough University, Loughborough LE11 3TU, UK  and Moscow State University, Moscow 119899, Russia}
\email{A.P.Veselov@lboro.ac.uk}

\begin{document}

\maketitle

\begin{abstract}
An infinite-dimensional version of Calogero-Moser operator of $BC$-type and the corresponding Jacobi symmetric functions are introduced and studied, including the analogues of Pieri formula and Okounkov's binomial formula. We use this to describe all the ideals linearly generated by the Jacobi symmetric functions and show that the deformed $BC(m,n)$ Calogero-Moser operators, introduced in our earlier work, appear here in a natural way as the restrictions of the $BC_{\infty}$ operator to the corresponding finite-dimensional subvarieties. As a corollary we have the integrability of these quantum systems and all the main formulas for the related super Jacobi polynomials. 
\end{abstract}

\tableofcontents

\section{Introduction} 
The classical Jacobi polynomials $P_n^{\alpha,\beta}(t)$ (sometimes called also as hypergeometric polynomials) are the orthogonal polynomials with respect to the weight $(1-t)^{\alpha}(1+t)^{\beta}$ on the interval $[-1,1].$ They satisfy the Jacobi differential equation
\begin{equation}
\label{Jac}
(1-t^2)y'' +[\beta-\alpha-(\alpha+\beta+2)t]y' + n(n+\alpha+\beta+1)y=0,
\end{equation}
and thus are particular case of the hypergeometric functions.

Replacing the weight by
$$\prod_{1\leq i<j \leq n}|t_i-t_j|^{2\theta}\prod_{1\leq i \leq n}(1-t_i)^{\alpha}(1+t_i)^{\beta}$$
on the $n$-dimensional cube we come to the multidimensional Jacobi polynomials $\mathcal J_{\lambda}(t; \theta,\alpha,\beta)$ (see e.g. \cite{OO}, where the parameters $\alpha, \beta$ are denoted by $a,b$). After change of coordinates and gauge transformation these polynomials are the eigenfunctions of the $BC_n$ version of {\it Calogero-Moser-Sutherland (CMS) operator} \cite{OP} and for special values of parameters can be interpreted as spherical functions on some compact symmetric spaces, including Grassmannians.

The $BC_n$ CMS operator is a particular case of the larger class of the so-called {\it deformed CMS operators} introduced in \cite{SV} in relation to generalised root system of type $BC(m,n)$ : 
\begin{eqnarray}
\label{bcnm} L^{m,n}& =& -\Delta_m
-k \Delta_n +\sum_{i<j}^{m}\left(\frac{2k(k+1)}{\sinh^2(x_{i}-x_{j})}+\frac{2k(k+1)}{\sinh^2(x_{i}+x_{j})}\right)\nonumber \\& &
+\sum_{i<j}^{n}\left(\frac{2(k^{-1}+1)}{\sinh^2(y_{i}-y_{j})}+\frac{2(k^{-1}+1)}{\sinh^2(y_{i}+y_{j})}\right)
\nonumber
\\& & +\sum_{i=1}^{m}\sum_{j=1}^{n}\left(\frac{2(k+1)}{\sin^2(x_{i}-y_{j})}+
\frac{2(k+1)}{\sinh^2(x_{i}+y_{j})}\right) +\sum_{i=1}^m
\frac{p(p+2q+1)}{\sinh^2x_{i}} \nonumber \\& & +\sum_{i=1}^m
\frac{4q(q+1)}{\sinh^22x_{i}} +\sum_{j=1}^n \frac{k
r(r+2s+1)}{\sinh^2y_{j}}+\sum_{j=1}^n \frac{4k
s(s+1)}{\sinh^22y_{j}},
\end{eqnarray}
where
$$\Delta_m=\left(\frac{\partial^2}{{\partial
x_{1}}^2}+\dots +\frac{\partial^2}{{\partial x_{m}}^2}\right), \,\, \Delta_n= \left(\frac{\partial^2}{{\partial y_{1}}^2}+ \dots
+\frac{\partial^2}{{\partial y_{n}}^2} \right),$$
and the parameters $k,p,q,r,s$ satisfy the
following relations
\begin{equation}
\label{rel}
p=kr,\quad 2q+1=k(2s+1).
\end{equation}
The usual $BC_m$ CMS operator corresponds to the case $n=0$, when there is no variables $y.$ 
The relation with the standard Jacobi parameters is
\begin{equation}
\label{relj}
\alpha=-p-q-\frac{1}{2}, \,\, \beta=-q-\frac{1}{2}
\end{equation}
and $\theta=-k/2$.
In the special case $n=1,\, p=r=0$ these operators were introduced in \cite{CFV1,CFV2, F1}, where the integrability of the corresponding quantum problems was established within the theory of Baker-Akhiezer function. In case when both $m$ and $n$ are larger than 1 this method can not be applied and the integrability was proved in \cite{SV} by a direct (but complicated) construction of the integrals.  In the case $A_{n,m}$ we gave in \cite{SV1}  a more conceptual proof of the integrability  by showing that the corresponding deformed CMS operator can be described as the restriction of the usual CMS operator with infinite number of variables onto certain subvarieties called generalised discriminants. 

The main goal of this paper is to extend this approach to the $BC$-case. This is not straightforward since in contrast to the $A$-case there is no stability for the CMS operators of $BC_n$ type. Our $BC_{\infty}$-version of  Calogero-Moser operator depends on an {\it additional parameter} $h$, which is related to  the dimension $N$. 
In the power sums coordinates $p_a$
it has the form
$$
 \mathcal{L}^{(k,p,q,h)}=\sum_{a,b>0}(p_{a+b}+2p_{a+b-1})\partial_{a}\partial_{b}-k\sum_{a=2}^{\infty}\left[\sum_{b=0}^{a-2}p_{a-b-1}(2p_{b}+p_{b+1})\right]\partial_{a}
$$
$$
+\sum_{a=1}^{\infty}\left[(a+k(a+1)+2h)p_{a}+(2a-1+2ka+2h-p)p_{a-1}\right]\partial_{a},
$$
where 
$$p_0=-k^{-1}(h+\frac{1}{2}p+q)$$ is the substitute for the dimension
and we use the notation $\partial_a = a \frac{\partial}{\partial p_a}$ (cf. \cite{Awata, Stanley}). This operator can be naturally represented as the sum of
the usual $A_{\infty}$ CMS operator with an additional $h$-dependent momentum term and the rational $BC_{\infty}$ Calogero-Moser operator (see Section 3).

We show that this $4$-parametric family  $\mathcal{L}^{(k,p,q,h)}$ has symmetry group $\mathbb Z_2^3$ generated by 3 commuting involutions (see Theorem \ref{symmetries} below)).
One of them is the extension of the duality $k \to k^{-1}$ well-known in $A_{\infty}$ case \cite{Ma}. It acts on other parameters as
$$2\hat h-1 = k^{-1}(2h-1),\,
\hat p=k^{-1}p, \,\, (2\hat q+1)=k^{-1}(2q+1),
$$
and thus gives an explanation of the relations (\ref{rel}) in the definition of the deformed CMS operator.
Note that we can not have this symmetry at the finite-dimensional level.

These involutions lead to the dualities for the {\it Jacobi symmetric functions} $\mathcal{J}_{\lambda}(u;k,p,q,h)$, which are defined as certain eigenfunctions of the operator $\mathcal{L}^{(k,p,q,h)}.$
These functions are the limiting case of the Koornwinder polynomials in infinitely many variables introduced by Rains \cite{Rains}.

We present the infinite dimensional analogues of Pieri formula, which in $BC_n$ case was found by van Diejen \cite{D}, and Okounkov's binomial formula, expressing Jacobi polynomials in terms of Jack polynomials \cite{Ok}. We use these formulas to describe all ideals, which are linearly spanned by the Jacobi symmetric functions ({\it $BC$-invariant ideals}) and show that the deformed CMS operator (\ref{bcnm}) after some gauge transformation is the restriction of the $BC_{\infty}$ operator onto the corresponding subvariety, which implies integrability of the corresponding quantum system
and a deformed version of the Harish-Chandra homomorphism.

The image of the Jacobi symmetric functions under this restriction gives the eigenfunctions of the corresponding commuting differential operators, which are called the {\it super Jacobi polynomials.}
We show that these polynomials have properties similar to Jacobi polynomials by presenting deformed versions of Pieri and Okounkov's formulas as well Opdam's evaluation formula \cite{Op}. Remarkably all these formulas can be written naturally in terms of the corresponding deformed root systems, which is another evidence of the importance of this notion. 

The Pieri formula can be interpreted as bispectrality between $BC(m,n)$ CMS operator (\ref{bcnm}) and a deformed version of the rational Koornwinder operator (see formulas (\ref{defKoorn}),(\ref{defKoornex}) below). In the special case $n=0$ and $n=1$ this bispectrality was established earlier by Chalykh \cite{Cha} and Feigin \cite{F2}, who used the technique of Baker-Akhiezer functions.

Although our initial motivation came from the theory of quantum integrable systems we believe that the most interesting applications of our results are in representation theory of orthosymplectic Lie superalgebras and related symmetric superspaces. We were also partly stimulated by an important paper by Okounkov and Olshanski \cite{OO}, where the asymptotic behaviour of the Jacobi polynomials $\mathcal J_{\lambda}$ as the number of variables and the Young diagram $\lambda$ go to infinity, was investigated.

\section{Jacobi polynomials  for the root system $BC_N$} 

Let $P_{N}={\mathbb C}[x^{\pm1}_{1},\dots,x^{\pm1}_{N}]$ be  the algebra of Laurent polynomials in $N$ independent variables  and  $W$ be the $BC_{N}$  Weyl group
$
W=S_{N}\ltimes\mathbb{Z}_{2}^{N}.
$ 
We have a natural action of the group  $W$ on  $P_{N}$: $S_{N}$ permutes the variables $ x_{1},\dots,x_{N}$ while  $\mathbb{Z}_{2}^{N}$  acts by $x_{i}\rightarrow x_{i}^{-1},\: i=1,\dots,\:N$. 
Let $P^W_{N} \subset P_N$
be the subalgebra  of the corresponding $W$-invariant Laurent polynomials; it is well-known that it can be identified with the standard algebra of symmetric polynomials (see below).

Recall that the generalised CMS operator related to any root system $R$ has the following form \cite{OP}
$$
L_{R}=-\Delta + \sum_{\alpha\in R_{+}}\frac{m_{\alpha}(m_{\alpha}+2m_{2\alpha}+1)(\alpha,\alpha)}{\sinh^2(\alpha,z)},\quad z \in {\mathbb R}^N,
$$
which is gauge equivalent to
$${\mathcal L}_R =-\Delta + 2\sum_{\alpha\in
R_{+}}m_{\alpha}\coth(\alpha, z)\partial_{\alpha},$$
where $\partial_{\alpha}=(\alpha, \frac{\partial}{\partial z})$ and $m_{\alpha}$ are some parameters ("multiplicities") prescribed to each root $\alpha$ in $W$-invariant way, $W$ is the corresponding Weyl group.
 In the case of $BC_{N}$ root system 
$$
R=\{\pm\varepsilon_{i}, \: \pm 2\varepsilon_{i}\: i=1,\dots, N,\: \pm\varepsilon_{i}\pm\varepsilon_{j},\: 1\le i <j \le N\},
$$
where $\varepsilon_{i}$ is the standard basis in $\mathbb R^{N},$
 we have the following operator
\begin{equation}
\label{CMS}
\begin{split}
{\mathcal L}^{(N)}=\sum_{i=1}^N (x_i \partial_{i})^2-&k\sum_{1\le i < j\le N} \left(\frac{x_{i}+x_{j}}{x_{i}-x_{j}}(x_i \partial_{i}-x_j \partial_{j})+ \frac{x_{i}x_{j}+1}{x_{i}x_{j}-1}(x_i \partial_{i}+x_j \partial_{j})\right) {}\\
& -\sum_{i=1}^N\left(p\frac{x_{i}+1}{x_{i}-1}+2q\frac{x^2_{i}+1}{x^2_{i}-1}\right)x_i \partial_{i},
\end{split}
\end{equation}
where $x_i = \exp 2 z_i, \, i=1,\dots, N$ and $k, p, q$ are the corresponding multiplicities
$$
m(\pm\varepsilon_{i})=p,\:m(\pm2\varepsilon_{i})=q,\: m(\pm\varepsilon_{i} \pm\varepsilon_{j})=k.
$$
Note that our multiplicities are different from the usual ones by a sign change.

The Jacobi polynomials $\mathcal P_{\lambda}(x,k,p,q)$ (in the Laurent form) are the eigenfunctions of this operator labelled by partitions $\lambda.$ 
Recall that
a {\it partition} is any sequence
$$
\lambda=(\lambda_{1},\lambda_{2},\dots,\lambda_{r}\,\dots)
$$
of nonnegative integers in decreasing order
$$
\lambda_{1}\ge\lambda_{2}\ge\dots\ge\lambda_{r}\,\ge\dots,
$$
which contains only finitely many nonzero terms. The number of nonzero terms is called the {\it length} of $\lambda$ and denoted by $l(\lambda).$ The sum  $$\mid\lambda\mid= \lambda_{1}+\lambda_{2}+\dots$$ is called the {\it weight} of $\lambda.$ 

In algebra $P_N^W$ there is the following natural analogue  of the  monomial symmetric functions:
$$
m_{\lambda}(x_{1},\dots,x_{N})= \sum x_{1}^{a_{1}}x_{2}^{a_{2}}\dots x_{N}^{a_{N}},
$$
where $\lambda$ is  a partition with $l(\lambda)\le N$ and the sum is taken  over all distinct sequences  $a_{1}, \dots, a_{n}$ such that   $|a_{1}|, \dots, |a_{n}|$ is a permutation of $\lambda$.
It is easy to see that  such $m_{\lambda}$  form a basis of $P_{N}^W$.

Consider the following partial order on the partitions. We write $\mu\le\lambda$ if and only if for all $ i=1,2, \dots$ we have
$$
\lambda_{1}+\dots+\lambda_{i}\le \mu_{1}+\dots+\mu_{i}.$$

The Jacobi polynomials $\mathcal P_{\lambda}(x,k,p,q),$ which usually are defined by the orthogonality condition (see e.g. \cite{OO}), can also be defined in the following way \cite{DLM,Heck}.

\begin{thm} 
\label{finjac}
If $1,k,p+2q$ are linearly independent over rational numbers  then for  any partition $\lambda$  , $l(\lambda)\le N$  there exists  a unique Laurent polynomial  $\mathcal  P_{\lambda}(x,k,p,q)\in P^W_{N}$  such that

1) $\mathcal  P_{\lambda}(x,k,p,q)= m_{\lambda}+\sum_{\mu<\lambda} u_{\lambda\mu}m_{\mu}$ for some $u_{\lambda\mu} = u_{\lambda\mu}(k,p,q)$

2) $ \mathcal  P_{\lambda}(x,k,p,q)$ is an eigenfunction of the operator ${\mathcal L}^{(N)}.$
\end{thm}

\begin{proof} 

Let $V_{\lambda}$ is a subspace in $P^W_{N}$ formed by linear combinations of $m_{\mu}$ where $\mu\le\lambda$.  It is known \cite{Heck}  that the operator ${\mathcal L}^{(N)}$
has an upper triangular matrix in the basis $m_{\mu}:$
$$
{\mathcal L}^{(N)}(m_{\lambda})=\sum_{\mu\le\lambda} c_{\lambda\mu}m_{\mu},
$$
where the coefficients $c_{\lambda\mu}$ can be described explicitly \cite{DLM}.
In particular, 
$$
c_{\lambda,\lambda}=\sum_{i=1}^N\lambda_{i}^2-2k\sum_{i=1}^N (N-i)\lambda_{i}-(p+2q)|\lambda|=
$$
$$
2n(\lambda^{\prime})+2k n(\lambda)-|\lambda|(p+2q-1+2k(N-1)),
$$
where $n(\lambda)=\sum_{i\ge 1}(i-1)\lambda_{i}$ and $\lambda^{\prime}$ is the conjugate partition (see e.g. \cite{Ma}). We have
$$
c_{\lambda,\lambda} - c_{\mu,\mu}=2k[n(\lambda)-n(\mu)]-2[n(\mu^{\prime})-n(\lambda^{\prime})]+(|\mu|- |\lambda|)(p+2q-1+2k(N-1)),
$$
which for generic $k, p, q$ is equal to 0 only if $|\lambda|=|\mu|$ and $n(\lambda)=n(\mu).$
But it is easy to check that if $|\lambda|=|\mu|$ and  $\mu\le\lambda$, then $n(\lambda)=n(\mu)$ only if $\lambda = \mu.$ This means that $ c_{\lambda,\lambda} \ne c_{\mu,\mu}$ if  $\lambda > \mu$ and the theorem follows now from linear algebra.
\end{proof} 
 
 In the new variables
\begin{equation}
\label{ui}
u_{i}=\frac12(x_{i}+x^{-1}_{i}-2), \, i=1,\dots, N
\end{equation}  the algebra 
$$P^W_N= \Lambda_N = {\mathbb C}[u_{1},\dots,u_{N}]^{S_N}$$ is the standard algebra of symmetric polynomials of $N$ variables and the Jacobi polynomials can be rewritten as symmetric polynomials
$$\mathcal P_{\lambda}(x_1,\dots, x_N; k,p,q) = \mathcal J_{\lambda} (u_1, \dots, u_N; k,p,q).$$ Slightly abusing  terminology we call $\mathcal J_{\lambda} (u_1, \dots, u_N; k,p,q)$ {\it Jacobi symmetric polynomials}. Note that because of the coefficient $1/2$ in formula (\ref{ui}) the coefficient at $m_{\lambda}$ in the expansion of  $\mathcal J_{\lambda}$ is $2^{|\lambda|}$ (but not 1). Note also that the value $x_i=1$ corresponds in the new variables to $u_i = 0.$

One can check that in the new coordinates  the operator (\ref{CMS}) has the following form 
\begin{equation}
\label{CMS1}
\begin{split}
 \mathcal{L}^{(N)}=&\sum_{i=1}^N u_i (u_i+2) \partial_{i}^2-k\sum_{ i<j}\frac{u_{i}+u_{j}}{u_{i}-u_{j}}\left(u_i\partial_{i}- u_j\partial_{j}\right) -4k\sum_{i<j}\frac{u_i\partial_{i}-u_j\partial_{j}}{u_{i}-u_{j}}
{}\\
&-(p+2q-1+k(N-1))\sum_{i=1}^N u_i\partial_{i}
-(2p+2q-1)\sum_{i=1}^N \partial_i.
\end{split}
\end{equation}
When $N=1$ this leads to the Jacobi differential equation (\ref{Jac}) with $$t=u_1+1.$$

It is interesting to note that the operator (\ref{CMS1}) can be represented as the sum
\begin{equation}
\label{CMS10}
\mathcal{L}^{(N)}=  \mathcal{A}^{(N)}+ \mathcal{B}^{(N)},
\end{equation}
where
\begin{equation}
\label{CMS11}
 \mathcal{A}^{(N)}=\sum_{i=1}^N (u_i\partial_{i})^2-k\sum_{i<j}^N\frac{u_{i}+u_{j}}{u_{i}-u_{j}}\left(u_i\partial_{i}-u_j\partial_{j}\right)-(p+2q+k(N-1))\sum_{i=1}^N u_i\partial_{i}
 \end{equation}
is the usual $A_{N-1}$ CMS operator (see e.g. formula (2) in \cite{SV1}) with an additional momentum term, 
and  \begin{equation}
\label{CMS12}
\mathcal{B}^{(N)}=2\sum_{i=1}^N u_i \partial_i^2 -4k\sum_{i<j}^N\frac{u_i\partial_{i}-u_j\partial_{j}}{u_{i}-u_{j}}-(2p+2q-1)\sum_{i=1}^N \partial_i,
  \end{equation}
which is the (gauged) rational $BC_N$ Calogero-Moser operator 
$$\mathcal L= \Delta-2k\sum_{i<j}^N\left(\frac{\partial_{i}-\partial_{j}}{x_{i}-x_{j}}+\frac{\partial_{i}+\partial_{j}}{x_{i}+x_{j}}\right)-2(p+q)\sum_{i=1}^N\frac{\partial_{i}}{x_{i}},\, \partial_i = \frac{\partial}{\partial x_i}$$
rewritten in coordinates $u_i=x_i^2.$

In the algebra  $\Lambda_N$ we have the usual basis consisting of monomial symmetric functions   
$$
m_{\lambda}(u_{1},\dots,u_{N})= \sum u_{1}^{a_{1}}u_{2}^{a_{2}}\dots u_{N}^{a_{N}},
$$
where the sum is taken  over all distinct sequences  $a_{1}, \dots, a_{n}$ such that   $a_{1}, \dots, a_{n}$ is a permutation of $\lambda$. It is easy to see that the transition matrix from the basis $\{m_{\lambda}(u_{1},\dots,u_{N})\}$  to the basis $\{m_{\lambda}(x_{1},\dots,x_{N})\}$ is upper triangular, so the operator $ \mathcal{L}^{(N)}$ has an upper triangular matrix in the basis $\{m_{\lambda}(u_{1},\dots,u_{N})\}$.

The following result due to Opdam \cite{Op} gives the value of the Jacobi polynomials $\mathcal  P_{\lambda}(1)=\mathcal  P_{\lambda}(1,\dots, 1;k,p,q)$ (which in the new coordinates is the same as $\mathcal J_{\lambda}(0)=\mathcal J_{\lambda}(0,\dots,0,k,p,q)$).

Let $$
R^{+}=\{\varepsilon_{i}, \:2\varepsilon_{i},\:\: \varepsilon_{i}\pm\varepsilon_{j},\: 1\le i <j \le N\}
$$ be the positive part of the $BC_N$ root system.  Introduce 
\begin{equation}
\label{rho}
\rho=\frac12\sum_{\alpha\in R^{+}}m_{\alpha}\alpha=\sum_{i=1}^{N}\left(k(N-i)+\frac12p+q\right)\varepsilon_{i}.
\end{equation}
and
\begin{equation}
\label{Upsilon}
\Upsilon(\lambda,\alpha)=\frac{\Gamma((\lambda-\rho,\alpha^{\vee})-m_{\alpha}-\frac12m_{\alpha/2}) \Gamma((-\rho,\alpha^{\vee})-\frac12m_{\alpha/2})}{\Gamma((\lambda-\rho,\alpha^{\vee})-\frac12m_{\alpha/2})\Gamma((-\rho,\alpha^{\vee})-m_{\alpha}-\frac12m_{\alpha/2})},
\end{equation}
where $\alpha^{\vee}= \frac{2 \alpha}{(\alpha, \alpha)}.$ For an explicit form of (\ref{Upsilon}) in the $BC_n$-case see Proposition \ref{combin}  below.

\begin{thm} (Opdam  \cite{Op}) 
\label{Opdam}
\begin{equation}
\label{j0}
\mathcal P_{\lambda} (1)=\prod_{\alpha\in R^{+}}\Upsilon(\lambda,\alpha).
\end{equation}
\end{thm}

 In order to define infinite dimensional analogs of Jacobi symmetric functions we need the following version of Opdam's formula (\ref{j0}) essentially due to Rains \cite{Rains}.
 
 Introduce for any partition $\lambda$ the following functions of one variable
\begin{equation}
\label{C+}
C^{+}_{\lambda}(x)=\prod_{(ij)\in\lambda}\left(\lambda_{i}+j+k(\lambda^{\prime}_{j}+i)+x\right),
\end{equation}
\begin{equation}
\label{C-}
C^{-}_{\lambda}(x)=\prod_{(ij)\in\lambda}\left(\lambda_{i}-j-k(\lambda^{\prime}_{j}-i)+x\right),
\end{equation}
\begin{equation}
\label{C0}
C^{0}_{\lambda}(x)=\prod_{(ij)\in\lambda}\left(j-1+k(i-1)+x\right).
\end{equation}

\begin{proposition}\label{combin} 
\begin{equation}
\label{j00}
\mathcal P_{\lambda} (1)=\mathcal J_{\lambda} (0)=4^{|\lambda|}\frac{C_{\lambda}^{0}(-kN)C_{\lambda}^{0}(k(1-N)-p-q+\frac12)}{C_{\lambda}^{-}(-k)C_{\lambda}^{+}(-2kN-p-2q-1)}.
\end{equation}
\end{proposition}

\begin{proof}
Let us rewrite the right hand side of the Opdam formula (\ref{j0}) as the product 
$$
\mathcal J_{\lambda} (0)=A^{+}A^{-}A^{0},
$$
where 
$$
A^{+}=\prod_{i< j}\frac{\Gamma(\lambda_{i}+\lambda_{j}+k(i+j-1-2N)-p-2q)}{\Gamma(\lambda_{i}+\lambda_{j}+k(i+j-2N)-p-2q)}
\prod_{i< j}\frac{\Gamma(k(i+j-2N)-p-2q)}{\Gamma(k(i+j-1-2N)-p-2q)}
$$
is the part corresponding to the roots $\varepsilon_{i}+\varepsilon_{j},\: i<j$,
$$
A^{-}=\prod_{1\le i< j\le N}\frac{\Gamma(\lambda_{i}-\lambda_{j}+k(i-j-1))}{\Gamma(\lambda_{i}-\lambda_{j}+k(i-j))}\prod_{1\le i< j\le N}\frac{\Gamma(k(i-j))}{\Gamma(k(i-j-1))}
$$
is the part corresponding to the roots $\varepsilon_{i}-\varepsilon_{j},\: i<j$ and
$$
A^{0}=\prod_{i=1}^N\frac{\Gamma(2\lambda_{i}+2k(i-N)-2p-2q)}{\Gamma(2\lambda_{i}+2k(i-N)-p-2q)}\prod_{i=1}^N\frac{\Gamma(2k(i-N)-p-2q)}{\Gamma(2k(i-N)-2p-2q)}
$$
$$
\prod_{i=1}^N\frac{\Gamma(\lambda_{i}+k(i-N)-p-2q)}{\Gamma(\lambda_{i}+k(i-N)-p-q)}\prod_{i=1}^N\frac{\Gamma(k(i-N)-p-q)}{\Gamma(k(i-N)-p-2q)}
$$
is the part corresponding to the roots $\varepsilon_{i}$ and $2\varepsilon_{i}.$

One can check that $C^{+}_{\lambda}(x)$ can be rewritten as
 $$
C^{+}_{\lambda}(x)=\prod_{1\le i\le j\le N}\left(\lambda_{i}+\lambda_{j+1}+k(i+j)+x+1\right)_{\lambda_{j}-\lambda_{j+1}}
$$
$$
=\prod_{i=1}^N\frac{\Gamma(2\lambda_{i}+2ki+x+1)}{\Gamma(\lambda_{i}+k(i+N)+x+1)}\prod_{1\le i< j\le N}\frac{\Gamma(\lambda_{i}+\lambda_{j}+k(i+j)+x+1)}{\Gamma(\lambda_{i}+\lambda_{j}+k(i+j-1)+x+1)},
$$
where we have used the standard notation $(x)_n = \frac{\Gamma(x+n)}{\Gamma(x)}.$
As a partial case for $\lambda=0$  we get an identity
$$
1=\prod_{1\le i< j\le N}\frac{\Gamma(k(i+j)+x+1)}{\Gamma(k(i+j-1)+x+1)}\prod_{i=1}^N\frac{\Gamma(2ki+x+1)}{\Gamma(k(i+N)+x+1)}.
$$
Using this identity we can rewrite the expression for $A^{+}$ as  follows
$$
A^{+}=\left[C_{\lambda}^{+}(x)\right]^{-1}\prod_{i=1}^N\frac{\Gamma(2\lambda_{i}+2ki+x+1)}{\Gamma(\lambda_{i}+k(i+N)+x+1)}\frac{\Gamma(k(i+N)+x+1)}{\Gamma(2ki+x+1)},
$$
where $x=-2kN-p-2q-1.$ 

In the same way we have
$$
A^{-}=\prod_{(ij)\in\lambda}\frac{j-1+k(i-1-N)}{\lambda_{i}-j-k(\lambda^{\prime}_{j}-i+1)}=\frac{C_{\lambda}^{0}(-kN)}{C_{\lambda}^{-}(-k)}
$$
and
$$
A^{0}=\prod_{i=1}^N\frac{\Gamma(2\lambda_{i}+2ki+x+1-p)}{\Gamma(2\lambda_{i}+2ki+x+1)}\frac{\Gamma(2ki+x+1)}{\Gamma(2ki+x+1-p)}
$$
$$
\prod_{i=1}^N\frac{\Gamma(\lambda_{i}+k(i+N)+x+1)}{\Gamma(\lambda_{i}+k(i+N)+x+q+1)}\prod_{i=1}^N\frac{\Gamma(k(i+N)+x+q+1)}{\Gamma(k(i+N)+x+1)}
$$
with the same $x=-2kN-p-2q-1.$ 

Therefore
$$
A^{+}A^{0}=(C_{\lambda}^{+}(x))^{-1}
\prod_{i=1}^N\frac{\Gamma(2\lambda_{i}+2ki+x+1-p)}{\Gamma(2ki+x+1-p)} 
$$
$$
\prod_{i=1}^N\frac{\Gamma(k(i+N)+x+q+1)}{\Gamma(\lambda_{i}+k(i+N)+x+q+1)}=  
4^{|\lambda|}\frac{C_{\lambda}^{0}(k(1-N)-p-q+1/2)}{C_{\lambda}^{+}(-2kN-p-2q-1)}.
$$
This implies the formula (\ref{j00}).
\end{proof}

Another important result is the following Pieri formula for Jacobi polynomials, which was found by van Diejen \cite{D}.

Let $J\subseteq I \subset \mathbb N$ be two finite subsets of natural numbers and $\sigma(J)=(\sigma_{j})_{j\in J}$ be a  sequence of signs $\sigma_{j} \in \{\pm 1\}, \, j\in J$. Define  the following functions of variables $x_{l}$, where $l\in I$:
$$
V^{+}_{\sigma(J),I}=\prod_{j\in J}w(\sigma_{j}x_{j})\prod_{i<j,\: i,j\in J}v(\sigma_{i}x_{i}+\sigma_{j}x_{j})v(\sigma_{i}x_{i}+\sigma_{j}x_{j}+1)
$$
$$
\prod_{j\in J,\; l\in I\setminus J}v(\sigma_{j}x_{j}+x_{l})v(\sigma_{j}x_{j}-x_{l})
$$

$$
V^{-}_{\sigma(J),I}=\prod_{j\in J}w(\sigma_{j}x_{j})\prod_{i<j,\: i,j\in J}v(\sigma_{i}x_{i}+\sigma_{j}x_{j})v(-\sigma_{i}x_{i}-\sigma_{j}x_{j}-1)
$$
$$
\prod_{j\in J,\; l\in I\setminus J}v(\sigma_{j}x_{j}+x_{l})v(\sigma_{j}x_{j}-x_{l}),
$$
where
$$
w(x)=\frac{(x-\frac12p-q)(x-\frac12p+\frac12)}{x(x+\frac12)},\quad v(x)=\frac{x-k}{x}
$$

 Let us assume from now on that the parameters $k, p, q$ are {\it generic} in the sense that together with 1 they are linearly independent over $\mathbb Q.$
  
 \begin{thm}\label{vand} (van Diejen \cite{D}) 
For generic values of parameters Jacobi symmetric polynomials satisfy the following Pieri identity 
 \begin{equation}
\label{pierifinite}
 2^rE_{r}\frac{\mathcal J_{\lambda}(u)}{\mathcal J_{\lambda}(0)}=\sum (-1)^{|L|}V^{-}_{\sigma(L),J^c}(\lambda-\rho)V^{+}_{\sigma(J),I}(\lambda-\rho)\frac{\mathcal J_{\lambda+\sigma(J)}(u)}{\mathcal J_{\lambda+\sigma(J)}(0)}
 \end{equation}
 where  $E_{r}$ is the  $r-th$ elementary symmetric function in variables $u_{1},\dots,u_{N}$, the sum is taken over  all $\sigma(L),\sigma(J)$, where $L, J \subseteq I= \{1,2,\dots,N\},\,L \cap J = \emptyset,\, |L|+|J|=r$ such that $\lambda+\sigma(J)$ is a partition, $J^c$ is the complement of $J$ in  $I$ and $\rho_{i}=kN+\frac12p+q-ki,\: i=1,\dots,N$.
\end{thm} 

For $r=1$ the Pieri formula can be rewritten as
 \begin{equation}
\label{pieri0}
 2 (\sum_{i=1}^N u_i )\frac{\mathcal J_{\lambda}(u)}{\mathcal J_{\lambda}(0)}=\sum_{\beta\in\mathcal O}V(\lambda,\beta)\left(\frac{\mathcal J_{\lambda+\beta}(u)}{\mathcal J_{\lambda+\beta}(0)}-\frac{\mathcal J_{\lambda}(u)}{\mathcal J_{\lambda}(0)}\right),
\end{equation}
where 
 \begin{equation}
\label{V0}
V(\lambda,\beta)=\prod_{\alpha\in R,(\alpha,\beta) > 0}\frac{\Upsilon(\lambda+\beta,\alpha)}{\Upsilon(\lambda,\alpha)},
 \end{equation}
 $\mathcal O=\{\pm \varepsilon_j|\,\, j=1,\dots,N\}$ and by $\lambda\pm\varepsilon_i$ we mean the partition $(\lambda_1, \dots, \lambda_i\pm 1, \dots, \lambda_N)$ when it is indeed a partition (otherwise the corresponding term is to be omitted in the right hand side). Geometrically this corresponds to the adding or deleting a box from the Young diagram.
In this form it can be generalised to the deformed case (see section 7 below).

In the algebra of symmetric functions there is another basis consisting of {\it Jack polynomials}
$P_{\lambda}(u,k)$ (see \cite {Ma}). A remarkable {\it binomial formula}
 due to Okounkov \cite{Ok, OO} expresses the Jacobi polynomials in terms of $P_{\lambda}(u,k)$. In order to write down this formula we need the following {\it interpolation $BC_N$ polynomials} $I_{\lambda}(z_{1},\dots,z_{N}; k,h)$  introduced and studied by Okounkov and Olshanski \cite{Ok,OO}. 
 
 Let $\Lambda^{(k,h)}_{N} \subset {\mathbb C}[z_{1},\dots, z_{N}]$ be the algebra of polynomials in $N$ variables $z_{1},\dots,z_{N},$ which are symmetric in variables $(z_{i}+ki + h)^2,\: i=1,\dots,N$, where $h$ is an additional parameter.
If 
\begin{equation}
\label{h}
h=-kN-\frac12p-q
\end{equation}
 then $ki +h = -\rho_i,$ where $\rho$ is given by (\ref{rho}), and the algebra $\Lambda^{(k,h)}_{N}$ can be interpreted  as the image under the Harish-Chandra homomorphism of the algebra $\mathcal D(N,k,p,q)$ of quantum integrals of the $BC_N$ Calogero-Moser system (see \cite{OP,Heck}). Namely, for any polynomial $f \in \Lambda^{(k,h)}_{N}$ with $h$ given by (\ref{h}) there exists a differential operator $\mathcal L_f$ commuting with (\ref{CMS1})
such that
\begin{equation}
\label{HC}
\mathcal L_f (\mathcal P_{\lambda} (x; k,p,q)) = f(\lambda) \mathcal P_{\lambda} (x; k,p,q).
\end{equation}

\begin{proposition} (Okounkov, Olshanski \cite{OO}) \label{intfinite}
For any $h$ and any partition $\lambda$  with $l(\lambda) \le N$ there exists a unique polynomial $I_{\lambda}(z_{1},\dots,z_{N}; k,h) \in \Lambda^{(k,h)}_{N}$  of degree $2|\lambda|,$ which satisfies the following conditions:
$$I_{\lambda}(\mu; k,h)=0$$ for any partition $\mu=(\mu_{1},\dots,\mu_{N})$ such that 
$\lambda\nsubseteq\mu,$ and normalisation condition
$$
I_{\lambda}(\lambda,k,h)=C_{\lambda}^-(1)C_{\lambda}^+(2h-1).
$$
\end{proposition}

Here we are using the usual geometric representation of the partitions as Young diagrams, so 
$\lambda\subseteq\mu$ means that the inclusion of the corresponding sets. 
For example, 
\begin{equation}
\label{I1}
I_{(1)}(z_{1},\dots,z_{N}; k,h)=\sum_{i=1}^N z_{i}(z_{i}+2ki+2h),
\end{equation}
where $(1)$ denotes the Young diagram consisting of only one box.

There exists the following combinatorial formula for these polynomials found by Okounkov \cite{Ok}:
\begin{equation}
\begin{split}
\label{comb}
I_{\lambda}(z_{1},\dots,z_{N}; k,h)=&\sum_{T}\psi_{T}(-k)\prod_{(i,j)\in\lambda}\left[(z_{T(i,j)}+h+kT(i,j))^2\right.
{}\\
&\left.-((j-1)+k(i-1)+h+kT(i,j))^2\right].
\end{split}
\end{equation}
Here the sum is taken over all reverse tableaux $T$ of shape $\lambda$ with entries in $\{1,\dots,N\}$, and  $\psi_{T}(-k)$ is the same weight as for ordinary Jack polynomials (see \cite{OO}). From this formula we see that the leading part of the polynomial $I_{\lambda}(z; k,h)$ coincides with
$P_{\lambda}(z_{1}^2,\dots,z_{N}^2;-k),$
where $P_{\lambda}(x,-k)$ is the corresponding Jack polynomial.

Now we are ready to present the binomial formula for Jacobi polynomials.

\begin{thm} (Okounkov \cite{Ok,OO})\label{binomial}
The Jacobi polynomials can be written in terms of Jack polynomials as follows:
\begin{equation}
\label{binom}
\mathcal J_{\lambda}(u_{1},\dots, u_{N}; k,p,q)=\sum_{\mu\subseteq\lambda}2^{|\mu|} \frac{\mathcal J_{\lambda}(0)I_{\mu}(\lambda; k,h)}{\mathcal J_{\mu}(0)I_{\mu}(\mu; k,h)}P_{\mu}(u_1,\dots,u_N; -k),
\end{equation}
where  $h=-kN-\frac12p-q$ and $l(\lambda) \le N$.
\end{thm}

As a corollary we have the following version of Pieri formula for interpolation $BC_N$ polynomials.
Let $a_{\lambda,\mu}$ be the matrix coefficients of the integrals $L_f$ in the basis of Jack polynomials:
$$
\mathcal L_{f}(P_{\lambda}(u,k))=\sum_{\mu\subset\lambda}a_{\lambda,\mu}P_{\mu}(u,k).
$$
They can be found from a triangular system of linear equations
$$
f(\lambda)C_{\lambda,\mu}=\sum_{\mu\subseteq\nu\subseteq\lambda}C_{\lambda,\nu}
a_{\nu,\mu}$$ with
$$C_{\lambda,\mu}= 2^{|\mu|} \frac{\mathcal P_{\lambda}(1)I_{\mu}(\lambda; k,h)}{\mathcal P_{\mu}(1)I_{\mu}(\mu;k,h)}.$$

\begin{proposition}\label{Pieri-int} For any $f\in\Lambda_N^{(k,h)}$ we have 
\begin{equation}
\label{pie}
f(\lambda) I_{\mu}(\lambda; k,h)=\sum_{\nu\supseteq\mu} C_{\mu,\nu} a_{\nu,\mu}I_{\nu}(\lambda; k,h),
\end{equation}
where $|\nu|-|\mu|\le \frac12deg(f)$ and the coefficients $C_{\mu,\nu}, \, a_{\lambda,\mu}$ are defined above.
\end{proposition}

\section{Jacobi symmetric functions of infinitely many variables}

 In this section we are going to generalise all the results from previous section to the algebra of symmetric functions in infinitely many variables.  Let $\Lambda$ be the projective limit of the graded algebras $\Lambda_N$ of symmetric polynomials in $N$ variables:
$$\Lambda = \lim_{\longleftarrow} \Lambda_{N}.$$ The power sums
$$p_l = u_1^l + u_2^l + \dots,\,\, l=1,2, \dots$$ is a convenient set of free generators of this algebra, the set
$$p_{\lambda}=p_{\lambda_1}p_{\lambda_2}\dots$$ for all partitions $\lambda$ forms a linear basis in $\Lambda.$
Consider a natural homomorphism $\varphi_{N}:\Lambda \longrightarrow \Lambda_{N}$
 defined by
\begin{equation}
\label{phin}
\varphi_{N}(p_{l})=\sum_{i=1}^N u^{l}_{i}.
\end{equation} 
An infinite dimensional version of the $BC_N$ CMS operator $\mathcal L^{(N)}$ is the operator $\mathcal L^{(k,p,q,h)}$, depending on extra parameter $h,$ replacing the dimension $N.$ To write it down 
it is convenient to introduce the notation
$$\partial_{a}=a\frac{\partial}{\partial p_{a}},\, a=1,2,\dots ..$$

 \begin{thm}\label{main} 
 {\em i)} There exists  a unique second order differential operator $\mathcal L^{(k,p,q,h)}: \Lambda \to \Lambda$ with coefficients, which are polynomial in all the parameters, such that for all $N=1,2,\dots$ and $h=-kN-\frac12p-q$ we have
\begin{equation}
\label{prop}
 \varphi_{N}\circ \mathcal L^{(k,p,q,h)}=\mathcal L^{(N)}\circ\varphi_{N},
 \end{equation}
where  $\mathcal L^{(N)}$ is the $BC_N$ CMS operator given by (\ref{CMS1}).
 
 {\em ii)} Operator $\mathcal L^{(k,p,q,h)}$ has the following explicit form in the power sums coordinates $p_{a}:$
 $$
 \mathcal{L}^{(k,p,q,h)}=\sum_{a,b>0}(p_{a+b}+2p_{a+b-1})\partial_{a}\partial_{b}-k\sum_{a=2}^{\infty}\left[\sum_{b=0}^{a-2}p_{a-b-1}(2p_{b}+p_{b+1})\right]\partial_{a}
$$
\begin{equation}\label{infinity}
+\sum_{a=1}^{\infty}\left[(a+k(a+1)+2h)p_{a}+(2a-1+2ka+2h-p)p_{a-1}\right]\partial_{a},
\end{equation}
with 
$$p_0=-k^{-1}(h+\frac{1}{2}p+q).$$
 \end{thm}
 
 \begin{proof} Represent the operator (\ref{infinity}) as the sum
 \begin{equation}
\label{suminf}
 \mathcal{L}^{(k,p,q,h)}= \mathcal{A}^{(k,h)}+ \mathcal{B}^{(k,p,q,h)}, 
 \end{equation}
 where
\begin{equation}
\label{CMS11inf}
\mathcal{A}^{(k,h)} = \sum_{a,b>0}p_{a+b}\partial_{a}\partial_{b}-k\sum_{a=2}^{\infty}\left(\sum_{b=0}^{a-2}p_{a-b-1}p_{b+1}\right)\partial_{a}+\sum_{a=1}^{\infty}\left[(a+k(a+1)+2h)p_{a}\right]\partial_{a},
 \end{equation}
\begin{equation}
\label{CMS12inf}
\mathcal{B}^{(k,p,q,h)}=2\sum_{a,b>0}p_{a+b-1}\partial_{a}\partial_{b}-2k\sum_{a=2}^{\infty}\left(\sum_{b=0}^{a-2}p_{a-b-1}p_{b}\right)\partial_{a}
\end{equation}
 $$+\sum_{a=1}^{\infty}(2a-1+2ka+2h-p)p_{a-1}\partial_{a},$$
 where $p_0=-k^{-1}(h+\frac{1}{2}p+q).$
 One can recognise in the first operator the usual $A_{\infty}$ CMS operator  (see \cite{Awata, Stanley}) with an additional $h$-dependent momentum term, while the second operator is actually the rational $BC_{\infty}$ Calogero-Moser operator (cf. formula (\ref{CMS10})).

To prove the theorem it is enough to show that  if $h = -kN - \frac12 p -q$  then for any $P \in \Lambda$
 $$
\mathcal{A}^{(N)}(\varphi_N(P))  = \varphi_N(\mathcal{A}^{(k,h)} (P))
 $$ 
and
$$
\mathcal{B}^{(N)}(\varphi_N(P))  = \varphi_N(\mathcal{B}^{(k,p,q,h)} (P)),
 $$
 where the operators  $\mathcal{A}^{(N)}$ and $\mathcal{B}^{(N)}$ are given by (\ref{CMS11}) and (\ref{CMS12}) respectively. 
Since the order of the corresponding differential operators is either one or two it is enough to prove this for power sums $P=p_a, a=1,2, \dots$ and their products $p_{a}p_{b}.$

One can check the following relations:
 $$
\mathcal{A}^{(N)}(p_{a})=(1+k)a^2p_{a}-ka\sum_{r=0}^{a-1}p_{r}p_{a-r}-a(p+2q+k(N-1))p_{a},
 $$
$$
\mathcal A^{(N)}(p_{a}p_{b})=\mathcal A^{(N)}(p_{a})p_{b}+p_{a}\mathcal A^{(N)}(p_{b})+2abp_{a+b},
 $$
  $$
\mathcal{B}^{(N)}(p_{a})=2(1+k)a^2p_{a-1}-2ka\sum_{r=0}^{a-1}p_{r}p_{a-r-1}-a(1+2p+2q)p_{a-1},
 $$
$$
\mathcal{B}^{(N)}(p_{a}p_{b})=\mathcal{B}^{(N)}(p_{a})p_{b}+p_{a}\mathcal{B}^{(N)}(p_{b})+4abp_{a+b-1}
 $$
with $p_0=N.$ The proof is straightforward with the use of the following identities: 
 $$
 \sum_{i < j}^N(u_{i}+u_{j})\frac{u_{i}^a-u_{j}^a}{u_{i}-u_{j}}=-ap_{a}+\sum_{i=0}^{a-1}p_{i}p_{a-i},
 $$
  $$
 \sum_{i < j}^N\frac{u_{i}^a-u_{j}^a}{u_{i}-u_{j}}=-\frac12ap_{a-1}+\frac12\sum_{i=0}^{a-1}p_{i}p_{a-i-1},
 $$
where again $p_0=N.$ The comparison with (\ref{CMS11inf}), (\ref{CMS12inf}) proves the existence.  The uniqueness part is evident.  \end{proof}

Consider now the following automorphisms  $\omega$ and $\theta$ of the algebra $\Lambda:$

\begin{equation}\label{omega}
\omega(p_{i})=k^{-1}p_{i},
\end{equation}
\begin{equation}\label{teta}
\theta(p_{i})=p_{i}+(-2)^{i}\frac{2k+1-2q}{2k}.
\end{equation}

\begin{thm}
\label{symmetries}
The $BC_{\infty}$-operator $\mathcal L^{(k,p,q,h)}$ has the following symmetries:
\begin{equation}\label{change2}
\mathcal L^{(k,p,q,h)}=\mathcal L^{(k,p^{\prime},q^{\prime},h)},
\end{equation}
\begin{equation}\label{changeom}
 \omega\circ\mathcal L^{(k,p,q,h)}=k \mathcal L^{(k^{-1},r,s,\hat h)}\circ\omega,
\end{equation}
\begin{equation}\label{change}
 \theta\circ\mathcal L^{(k,p,q,h)}=\mathcal L^{(k,\tilde p,\tilde q,h)}\circ\theta,
\end{equation}
where 
\begin{equation}\label{sym1}
 p^{\prime}=1+2k-p-2q,\:q^{\prime}=q,
\end{equation}
\begin{equation}\label{symom}
2\hat h-1 = k^{-1}(2h-1),\,\, r=k^{-1}p, \,\, (2s+1)=k^{-1}(2q+1),
\end{equation}
\begin{equation}\label{sym2}
\tilde p=-p,\: \tilde q=2k+1-q.
\end{equation}
\end{thm}

\begin{proof} The symmetries (\ref{change2}),(\ref{changeom}) hold for $\mathcal{A}^{(k,h)}$ and $\mathcal{B}^{(k,p,q,h)}$ separately and  are obvious from their form. The third symmetry holds only for the full operator (\ref{infinity}). 
Let us look more generally for the shift symmetry of the form
$$p_a \rightarrow p_a +\gamma_a.$$
From the second order part we see immediately that constants $\gamma_a$ must satisfy the relation
$\gamma_{a+1}+2\gamma_a=0,$
which implies that 
$\gamma_a = (-2)^a \gamma$
for some constant $\gamma.$ A simple analysis shows that this gives the symmetry of the operator 
(\ref{infinity}) if and only if
$$\gamma= \frac{2k+1-2q}{2k}$$
and the corresponding parameters are related by the formula (\ref{sym2}) or by the formula
$$\tilde p = p+2q-2k-1, \tilde q=2k+1-q,$$ which corresponds to the composition of $\theta$ with the first symmetry 
(\ref{sym1}).
\end{proof}

{\bf Remark.} The symmetry (\ref{changeom}) is an analogue of duality $k \to k^{-1}$ in the theory of Jack polynomials (see \cite{Ma}). The fact that the corresponding change of parameters (\ref{symom}) coincides with the relations (\ref{rel}) in the definition of the deformed $BC(m,n)$ operators (similarly to the $A(m,n)$-case \cite{SV1}) seems to be important and shed more light on the nature of the deformed root systems \cite{SV}. We note also that one can not see any of these symmetries at the finite-dimensional level.

Let us define now the {\it Jacobi symmetric functions of infinitely many variables} $\mathcal J_{\lambda}(u; k,p,q,h) \in \Lambda$.

\begin{thm} 
If $1,k,h$ are linearly independent over rational numbers  then for  any partition $\lambda$  there exists  a unique  polynomial  $\mathcal  J_{\lambda}(u; k,p,q,h)\in\Lambda$  such that

1) $\mathcal  J_{\lambda}(u; k,p,q,h)= 2^{|\lambda|} m_{\lambda}(u)+\sum_{\mu<\lambda} u_{\lambda\mu}m_{\mu}(u)$ for some $u_{\lambda\mu} = u_{\lambda\mu}(k,p,q,h)$

2) $ \mathcal  J_{\lambda}(u; k,p,q,h)$ is an eigenfunction of ${\mathcal L}^{(k,p,q,h)}$ with the eigenvalue
$$
2n(\lambda^{\prime})+2k n(\lambda)+|\lambda|(2h+2k+1),
$$
where as before $n(\lambda)=\sum_{i\ge 1}(i-1)\lambda_{i}$ and $\lambda^{\prime}$ is the conjugate partition. 
\end{thm}

The proof is the same as in the finite-dimensional case (see Theorem \ref{finjac}) and uses the upper-triangularity of the operator in the monomial basis.

{\bf Remark.} As it follows from Okounkov's binomial formula (see (\ref{jinf}) below) the Jacobi symmetric functions $J_{\lambda}(u; k,p,q,h))$ are well-defined under slightly weaker conditions, namely when $k$ is not a positive rational and $$h \neq \frac{1}{2}(ak+b)$$ for all integers $a$ and $b.$
 
\begin{proposition}\label{parameters} The Jacobi symmetric functions have the following symmetries:
 \begin{equation}\label{change3} 
\mathcal J_{\lambda}(u; k,p,q,h))=\mathcal J_{\lambda}(u; k, p^{\prime}, q^{\prime},h),
 \end{equation}
 \begin{equation}\label{change31}
 \omega (\mathcal J_{\lambda}(u; k,p,q,h))=\frac{\mathcal J_{\lambda}(0; k,p,q,h)}{\mathcal J_{\lambda'}(0; k^{-1},r,s,\hat h)}\mathcal J_{\lambda'}(u; k^{-1},r,s,\hat h),
 \end{equation}
 \begin{equation}\label{change1}
 \theta(\mathcal J_{\lambda}(u; k,p,q,h))=\mathcal J_{\lambda}(u; k,\tilde p,\tilde q,h),
 \end{equation}
 where $\lambda'$ is conjugate to $\lambda$ and all parameters are the same as in Theorem \ref{symmetries}.
\end{proposition}

\begin{proof}
The first two symmetries follow from the previous theorem and formulas (\ref{change2}), (\ref{changeom}).
To prove relation (\ref{change1}) it is enough to show that the automorphism (\ref{teta}) has an upper-triangular form in the monomial basis.

We prove this for the following more general automorphism $\theta_{q,r}$ defined by
\begin{equation}\label{teta1} 
\theta_{q,r}(p_{i})=p_{i}+rq^{i},
 \end{equation}
 where $r$ and $q$ are arbitrary parameters. 
For a given partition $\lambda$ consider the set of all different partitions $\mu$, which one can get from $\lambda$ by eliminating at most one part of it (or one row in the Young diagram representation).
We have the following obvious formula:
\begin{equation}\label{mon}
m_{\lambda}(x_{1},\dots,x_{n},q)=\sum_{\mu\cup(a)=\lambda}q^{a}m_{\mu}(x_{1},\dots,x_{n}),
\end{equation}
where  $(a)$ denote the row of length $a.$  Iterating this formula $N$ times we get
\begin{equation}\label{mon1}
m_{\lambda}(x_{1},x_{2},\dots,x_{n},q,\dots,q)=\sum_{\mu}q^{|\lambda|-|\mu|}N_{\lambda,\mu}m_{\mu}(x_{1},\dots,x_{n}),
\end{equation}
where $\mu$ runs over all partitions, which can be obtained  from $\lambda$ by removing some of its rows, and  $N_{\lambda,\mu}$ is the number of  sequences of nonnegative integers $a_{1},\dots,a_{N}$ such that $\mu\cup(a_{1})\cup\dots\cup(a_{N})=\lambda.$

Now we note that if $r$ is a positive  integer
then $$\theta_{q,r}(m_{\lambda})=m_{\lambda}(x,q,\dots,q),$$
 where $q$ occurs $r$ times. 
The triangularity of $\theta_{q,r}$ in that case follows from (\ref{mon1}). Since the coefficients are polynomial in $r$ this is true for all $r,$ which completes the proof.
\end{proof}

To define higher order integrals of $BC_{\infty}$ CMS system we need the following infinite dimensional version of the shifted $BC_N$ polynomials \cite{OO}.
Consider  the projective limit
  $$
 \Lambda^{(k,h)}=\lim_{\longleftarrow} \Lambda^{(k,h)}_{N}
 $$
 taken in the category of filtered algebras with respect to the homomorphisms sending the last variable to $0$ and the filtration defined by the total degree of polynomial.
By definition any element $f$ from $\Lambda^{(k,h)}$ is a sequence of elements $f_{N}\in \Lambda^{(k,h)}_{N}$ such that $f_{N+1}(z_{1},\dots,z_{N},0)=f_{N}(z_{1},\dots,z_{N})$ and the degrees of polynomials $f_{N}$ are bounded. We can evaluate element $f\in\Lambda^{(k,h)}_{N}$ at any infinite vector $z=(z_{1},z_{2},\dots)$ with finitely many nonzero coordinates.

\begin{proposition} (Okounkov, Olshanski \cite{OO}) \label{intinf} 
For any partition $\lambda$ there exists  $I_{\lambda} \in \Lambda^{(k,h)}$  of degree $deg( I_{\lambda})=2|\lambda|$, which satisfies the conditions 
$$I_{\lambda}(\mu,k,h)=0$$ for any partition $\mu$ such that 
$\lambda\nsubseteq\mu,$ and the normalisation condition
$$
I_{\lambda}(\lambda,k,h)= C^{-}_{\lambda}(1) C^{+}_{\lambda}(2h-1).
$$
\begin{proof}
Consider the sequence of polynomials 
$
\{I_{\lambda}(z_{1},\dots,z_{N}, k,h)\}_{N\ge l(\lambda)}.
$
From formula (\ref{comb}) we have 
$$
I_{\lambda}(z_{1},\dots,z_{N},0, k,h)=I_{\lambda}(z_{1},\dots,z_{N}, k,h),
$$
so this sequence defines an element from $I_{\lambda}(k,h)$ with required properties.\end{proof}
\end{proposition}

The following theorem gives an infinite-dimensional version of the Harish-Chandra homomorphism.

 \begin{thm}
 \label{HaCh}
For any $f\in \Lambda^{(k,h)}$ there exists a unique differential operator $\mathcal L^{(k,p,q,h)}_{f}$ on $\Lambda$ such that for any positive integer $N$ the following diagram is commutative
 $$
\begin{array}{ccc}
\Lambda&\stackrel{\mathcal L^{(k,p,q,h)}_{f}}{\longrightarrow}& \Lambda\\
\downarrow \lefteqn{\varphi_{N}}& &\downarrow \lefteqn{\varphi_{N}}\\
\Lambda_{N}&\stackrel{\mathcal L^{(N)}_{f}}{\longrightarrow}&\Lambda_{N}
 \\
\end{array}
$$ 
where $h=-kN-\frac12 p -q,$ $\varphi_{N}$ is defined by (\ref{phin}) and the differential operators $\mathcal L^{(N)}_{f}$ are defined by the Harish-Chandra homomorphism (\ref{HC}).
\end{thm} 

\begin{proof}
Define the operator $ \mathcal L^{(k,p,q,h)}_{f}$ by its action on the Jacobi symmetric functions:  
$$\mathcal L^{(k,p,q,h)}_{f}(\mathcal J_{\lambda} (u; k,p,q,h))=f(\lambda)\mathcal J_{\lambda} (u; k,p,q,h).
$$ 
 If  $h=-kN-\frac12 p -q$  then from Theorem \ref{main} we see that 
 $$
 \varphi_N ( \mathcal J_{\lambda} (u; k,p,q,h))=\mathcal J_{\lambda}(u_{1},\dots,u_{N},k,p,q)
 $$
 
 Therefore  the operator $\mathcal L^{(k,p,q,h)}_{f}$ reduces to the operator $\mathcal L^{(N)}_{f}$ in $\Lambda_N$ with the property 
$$L^{(N)}_f (\mathcal J_{\lambda} (u_1,\dots, u_N; k,p,q)) = f(\lambda) \mathcal J_{\lambda} (u_1, \dots, u_N; k,p,q).$$ By Harish-Chandra homomorphism (\ref{HC}) $\mathcal L^{(N)}_{f}$ is a differential operator of order equal to $deg(f)$.

Let us prove  now that  $ \mathcal L_{f}$ is also a differential operator of the same order $l=deg(f)$. By definition it is enough to prove that 
$$
\left[\left[\left[ \mathcal L_{f},g_{1}\right], g_{2}\right]\dots,g_{l+1}\right]=0
$$
for all $g_1, \dots, g_{l+1} \in \Lambda.$ Fix $g_1, \dots, g_{l+1}$ and denote $\left[\left[\left[ \mathcal L_{f},g_{1}\right], g_{2}\right]\dots,g_{l+1}\right]$ as $D=D[h] \in \Lambda.$ Express
 $$
D = \sum_{\lambda} c_{\lambda}(h)m_{\lambda}
$$
 as a finite linear combination of monomial functions $m_{\lambda}.$ Let $M$ be the maximum of the lengths of all the corresponding partitions $\lambda.$ Since for all $N>M$ the images $\varphi_{N}(m_{\lambda})$
are linearly independent and $\varphi_N(D)=0$ for $h=-kN-\frac12 p - q$, the corresponding coefficients 
must be equal to 0 for infinitely many values of $h$. Since they are rational functions of $h$ they must be equal identically to zero.
\end{proof}

\section{Pieri formula and $BC$-invariant ideals}

In this section we describe all the ideals $I \subset \Lambda,$ which have a linear basis consisting of Jacobi symmetric functions. Since this is equivalent to the invariance of the ideal $I$ with respect to all quantum integrals $\mathcal L_f^{(k,p,q,h)}$ of $BC_{\infty}$ CMS system we will call such ideals $BC$-{\it invariant} (cf. \cite{SV1}).

We start with an infinite dimensional analog of the Pieri formula, which in the $BC_N$ case was found by van Diejen \cite{D}.  We should mention that in the form (\ref{pierifinite}) Pieri formula contains many zero summands.  The following lemma shows that the number of the non-zero coefficients in (\ref{pierifinite}) for large $N$ depend only on the diagram $\lambda$ (but not on $N$).

\begin{lemma}\label{zero} 
Let $L,\,J, \sigma$ be the same as in Theorem \ref{vand}
and $M(L)$ and $M(J)$ be  the maximal elements of $L$ and $J$ respectively. If $\lambda+\sigma(J)$ is a partition and  $M(L)\ge l(\lambda)+|L|+1,\: M(L)\ge M(J)+|L|+1$ then $V^{-}_{\sigma(L),J^c}(\lambda-\rho)=0$.

\begin{proof}
Let $l=M(L)$  and $p$ be  the maximal  nonnegative integer 
such that $\{l,l-1,\dots, l-p\}\subset L$.  By assumption  we have $l-p-1\ge M(J)+|L|-p>M(J)$, so  $l+1\notin J$ and $l-p-1\notin J$. This means that $V^{-}_{\sigma(L),J^c}(\lambda-\rho)$ with $\rho$ given by (\ref{rho}) contains a factor
$$
A=v(\sigma_{l}x_{l}+x_{l+1})v(\sigma_{l-p}x_{l-p}-x_{l-p-1})\prod_{i=l-p-1}^{l}v(\sigma_{i-1}x_{i-1}+\sigma_{i}x_{i})
$$
where $x_{i}=\lambda_{i}+ki+h,\: i=1,\dots,N$ and $h=-kN-\frac12p-q.$  Let us show that for any choice of $\sigma_{r},\: r\in L$ the product $A$ is equal to $0$.
By assumptions we have  $l-p-1> l(\lambda)$, therefore $x_{l-p}= k(l-p)+h,\: x_{l-p-1}= k(l-p-1)+h$. So if $\sigma_{l-p}=1$, then $v(\sigma_{l-p}x_{l-p}-x_{l-p-1})=v(k)=0$. The same arguments show that if $\sigma_{l}=-1$ then  $v(\sigma_{l}x_{l}+x_{l+1})=0$. Therefore we can assume that $\sigma_{l-p}=-1,\sigma_{l}=+1$. But in this case  there exist $i$ such that   $l-p+1\ge i\ge l$ and $\sigma_{i-1}=-1,\: \sigma_{i}=1$ and $v(\sigma_{i-1}x_{i-1}+\sigma_{i}x_{i})=0$. 
 \end{proof}
\end{lemma}

Note that the inequalities of Lemma \ref{zero} are automatically satisfied if
$$M(L) \geq l(\lambda) + 2r +1,$$
where $r=|L|+|J|.$

Now let us fix  a natural number $r$, a partition $\lambda$ and two disjoint finite sets $L, J \subset \mathbb N$ such that $\lambda+\sigma(J)$ is a partition, $M(L) < l(\lambda) + 2r +1$  and $|L|+|J|=r.$ Let us fix also the set of signs $\sigma(L),\sigma(J)$. Note that if $N\ge l(\lambda) + 2r +1$ both sets $L$ and $J$ belong to $I = \{1,\dots, N\}.$

\begin {lemma}\label{infty}
There exist unique rational functions $V^-_{\sigma(L),J,\lambda}(h),\:V^+_{\sigma(J),\lambda}(h)$  such that  for any set $I=\{1,\dots,N\}$ with $N\ge l(\lambda) + 2r +1$ we have 
$$
V^-_{\sigma(L),J,\lambda}(-kN-\frac12p-q)=V^-_{\sigma(L),J^c}(\lambda-\rho),
$$ 
$$
V^+_{\sigma(J),\lambda}(-kN-\frac12p-q)=V^+_{\sigma(J),I}(\lambda-\rho).
$$
The rational functions $V^{+}_{\sigma(J),\lambda}(h),\:V^{-}_{\sigma(L),J,\lambda}(h)$ have the following form
$$
V^{+}_{\sigma(J),\lambda}(h)=Q_{\lambda}^+(h)\prod_{j\in J}(\lambda_{j}+kj+h-\frac12\sigma_{j}(p-1))(\lambda_{j}+kj+h+\sigma_{j}(\frac12p+q-k)),
$$
$$
\quad V^{-}_{\sigma(L),J\lambda}(h)=Q_{\lambda}^-(h)\prod_{l\in L}(\lambda_{l}+kl+h-\frac12\sigma_{l}(p-1))(\lambda_{l}+kl+h+\sigma_{l}(\frac12p+q-k)),
$$
where $Q_{\lambda}^+(h),Q_{\lambda}^-(h)$ are some rational functions of $h$ with  poles and zeros of the form
$\frac12(a+kb)$ with some integers $a,b$. 
\end{lemma}

We also need the following

\begin{proposition} 
\label{jzero}
The value of the Jacobi symmetric functions at zero is
\begin{equation}
\label{jinf0}
\mathcal J_{\lambda}(0)=4^{|\lambda|}\frac{C_{\lambda}^{0}(h+\frac12 p + q)C_{\lambda}^{0}(k+h-\frac12 p +\frac12)}{C_{\lambda}^{-}(-k)C_{\lambda}^{+}(2h-1)}.
\end{equation}
\end{proposition}

The proof follows from the finite-dimensional formula (\ref{j00}).

Now we are ready to give an infinite dimensional version of the Pieri formula.
Recall that the parameters $p,q.k$ are generic if  $p,q,k$ and 1 are linearly independent over $\mathbb Q$. We say also that $h$ is {\it admissible} if $$h \neq \frac{1}{2}(ak+b)$$ for all non-positive integers $a$ and $b.$ We know that under these conditions Jacobi symmetric functions are well-defined.

\begin{thm}\label{Pieriinfty}
For generic $p,q, k$ and admissible $h$ Jacobi symmetric functions satisfy the following Pieri identity 
 \begin{equation}
\label{pieriinfty}
 2^rE_{r}\mathcal J_{\lambda}=\sum (-1)^{|L|}V^{-}_{\sigma(L),J,\lambda}(h)V^{+}_{\sigma(J),\lambda}(h) \frac{\mathcal J_{\lambda}(0)}{\mathcal J_{\lambda+\sigma(J)}(0)} \mathcal J_{\lambda+\sigma(J)},
 \end{equation}
where  $E_{r}$ is the $r$-th elementary symmetric function and the sum is taken over all $\sigma(L),\sigma(J)$ with $L,J \subset \mathbb N$ being the disjoint finite sets such that $|L|+|J|=r,\, M(L) \le l(\lambda) + 2r +1$ and $\lambda+\sigma(J)$ is a partition.
\end{thm}

The proof follows from lemmas \ref{zero}, \ref{infty} and formula (\ref{jinf0}).  Note that all coefficients have a natural regularisation for all admissible $h$ due to some cancellations of the common factors in the numerators and denominators.

In the case $r=1$ we can make the formula more explicit.

\begin{corollary} \label{pieri1} 
Jacobi symmetric functions satisfy the following Pieri formula 
\begin{equation}
\label{pieriinf}
 2 p_{1}(u)\mathcal J_{\lambda}(u)=\sum_{\beta}V(\lambda,h,\beta)\left(\frac{\mathcal J_{\lambda}(0)}{\mathcal J_{\lambda+\beta}(0)}\mathcal J_{\lambda + \beta}(u) - \mathcal J_{\lambda}(u)\right),
\end{equation}
where $E_1=p_1$ is the first power sum,
$$
 V(\lambda,h,\varepsilon_i)= \prod_{ j\ne i}^{l(\lambda)+1}\frac{(\lambda_{i}-\lambda_{j}+k(i-j-1))(\lambda_{i}+\lambda_{j}+k(i+j-1)+2h)}{(\lambda_{i}-\lambda_{j}+k(i-j))(\lambda_{i}+\lambda_{j}+k(i+j)+2h)}
$$
 $$
 \frac{(\lambda_{i}+k(i-1) + h + \frac12 p +q)(\lambda_{i}+k(i+l(\lambda)+1)+2h)(\lambda_{i}+ki +h - \frac12 p + \frac12)}{(\lambda_{i}+k(i-l(\lambda)-2))(\lambda_i + ki +h)(\lambda_i + ki +h + \frac12)},
 $$
 $$
 V(\lambda,h, -\varepsilon_i)= \prod_{ j\ne i}^{l(\lambda)}\frac{(\lambda_{i}-\lambda_{j}+k(i-j+1))(\lambda_{i}+\lambda_{j}+k(i+j+1)+2h)}{(\lambda_{i}-\lambda_{j}+k(i-j))(\lambda_{i}+\lambda_{j}+k(i+j)+2h)}
$$
 $$
 \frac{(\lambda_{i}+k(i+1)+h-\frac12 p -q)(\lambda_{i}+k(i-l(\lambda)))(\lambda_{i}+ki +h + \frac12 p - \frac12)}{(\lambda_{i}+k(i+l(\lambda)+1)+2h)(\lambda_i + ki +h)(\lambda_i + ki +h - \frac12)},
 $$
and the sum is taken over all  $\beta = \pm \varepsilon_i, \, i=1,2, \dots$ such that $\lambda + \beta$ is a Young diagram.
\end{corollary}
 
We can rewrite it in combinatorial way  similar to Macdonald  (see [\cite{Ma},VI.6]) using the following "content" functions (cf. (\ref{C+})-(\ref{C0})):
\begin{equation}
\label{c+}
c^{+}_{\lambda}(\Box, x)= \lambda_{i}+j+k(\lambda^{\prime}_{j}+i)+x,
\end{equation}
\begin{equation}
\label{c-}
c^{-}_{\lambda}(\Box, x)= \lambda_{i}-j-k(\lambda^{\prime}_{j}-i)+x,
\end{equation}
\begin{equation}
\label{c0}
c^{0}_{\lambda}(\Box, x)= j-1+k(i-1)+x.
\end{equation}
Let $\mu=\lambda+\varepsilon_i$ be the Young diagram, which is the result of adding to $\lambda$ one box in $i$-th row. Consider the rectangle $(l(\lambda)+1)\times(l(\lambda')+1)$ and the vertical strip $\pi_i$, consisting of boxes in this rectangle belonging to the same column as added box (excluding the last one).

Then the coefficients in Pieri formula (\ref{pieriinf}) can be rewritten as
\begin{equation}
\label{V+}
V(\lambda,h,\varepsilon_i)= K_{\lambda}^+(\boxdot,h) \prod_{\Box\in\pi_i} B_{\lambda}^+(\Box,h),
\end{equation}
where 
\begin{equation}
\label{b+}
B_{\lambda}^+(\Box,h)=\frac{c_{\lambda}^{-}(\Box, 1)c_{\lambda}^{+}(\Box, 2h-1)}{c_{\mu}^{-}(\Box, 1)c_{\mu}^{+}(\Box, 2h-1)},
\end{equation}
\begin{equation}
\label{Ki+}
K_{\lambda}^+ (\boxdot,h)= \frac{c_{\mu}^{0}(\boxdot, h+\frac{1}{2}p + q))c_{\mu}^{0}(\boxdot, k(l(\lambda)+2)+2h) c_{\mu}^{0}(\boxdot, h-\frac{1}{2}p + \frac{1}{2} +k)}{c_{\mu}^{0}(\boxdot, h+k))c_{\mu}^{0}(\boxdot, -k(l(\lambda)+1) c_{\mu}^{0}(\boxdot, h+ \frac{1}{2} +k)}
\end{equation}
and $\boxdot$ denotes the added box in the $i$-th row.

Similarly,
\begin{equation}
\label{V-}
V(\lambda,h,-\varepsilon_i)= K_{\lambda}^-(\boxdot,h) \prod_{\Box\in\pi_i} B_{\lambda}^-(\Box,h),
\end{equation}
where 
\begin{equation}
\label{b-}
B_{\lambda}^-(\Box,h)=
\prod_{\Box\in\pi_i}\frac{c_{\mu}^{-}(\Box, 2k)c_{\mu}^{+}(\Box, 2k+2h)}{c_{\lambda}^{-}(\Box, 0)c_{\lambda}^{+}(\Box, 2h)},
\end{equation}
\begin{equation}
\label{Ki-}
K_{\lambda}^-(\boxdot,h) = \frac{c_{\lambda}^{0}(\boxdot, 2k+1+h-\frac{1}{2}p - q))c_{\lambda}^{0}(\boxdot, 1+k(1-l(\lambda)) )c_{\lambda}^{0}(\boxdot, k+h+\frac{1}{2}p + \frac{1}{2})}{c_{\lambda}^{0}(\boxdot, 1+k+h)c_{\lambda}^{0}(\boxdot, 1+k(l(\lambda)+2)+2h) c_{\lambda}^{0}(\boxdot, h+ \frac{1}{2} +k)},
\end{equation}
 $\mu=\lambda-\varepsilon_i$ and $\boxdot$ denotes the box taken out from the $i$-th row.

Comparing with Macdonald's form of Pieri formula for Jack polynomials (see formulas (6.24) in \cite{Ma}) we see that the vertical strip $\pi_i$ goes beyond the diagram $\lambda$ within some rectangle containing this diagram. The choice of this rectangle is actually not very relevant: one can check that for any rectangle $M\times N$ containing the Young diagram $\mu = \lambda+ \varepsilon_i$ (in particular, with $M>l(\lambda), \, N > l(\lambda')$) we have the same formulas (\ref{V+}),(\ref{V-}) with
\begin{equation}
\label{Ki+MN}
K_{\lambda}^+ (\boxdot,h) = \frac{c_{\mu}^{0}(\boxdot, h+\frac{1}{2}p + q))c_{\mu}^{0}(\boxdot, k(M+1)+2h) c_{\mu}^{0}(\boxdot, h-\frac{1}{2}p + \frac{1}{2} +k)}{c_{\mu}^{0}(\boxdot, h+k))c_{\mu}^{0}(\boxdot, -kM) c_{\mu}^{0}(\boxdot, h+ \frac{1}{2} +k)},
\end{equation}
\begin{equation}
\label{Ki-MN}
K_{\lambda}^- (\boxdot,h) = \frac{c_{\lambda}^{0}(\boxdot, 2k+1+h-\frac{1}{2}p - q))c_{\lambda}^{0}(\boxdot, 1+k(1-M) )c_{\lambda}^{0}(\boxdot, k+h+\frac{1}{2}p + \frac{1}{2})}{c_{\lambda}^{0}(\boxdot, 1+k+h)c_{\lambda}^{0}(\boxdot, 1+k(M+2)+2h) c_{\lambda}^{0}(\boxdot, h+ \frac{1}{2} +k)}.
\end{equation}

Now we are ready to describe all $BC$-invariant ideals $I \subset \Lambda.$ It turns out that in contrast to the $A_n$-case \cite{SV1} these ideals admit a simple description.

\begin{thm}\label{singular} Let parameters $p,q,k$ be  generic and $h$ be admissible. 
A non-trivial $BC$-invariant ideal in $\Lambda$ exists if and only if
\begin{equation}
\label{1}
h= -km-n - \frac12p-q,\: h=-k(m+1)-n - \frac12 p+\frac12 
\end{equation}
or
\begin{equation}
\label{2}
h=-k(m+2)-(n+1) + \frac12p+q,\: h=-k(m+1)-n + \frac12 p-\frac12
\end{equation}
for some non-negative integers $m$ and $n,$ in which case it is unique. 

The corresponding ideal is  linearly generated by the Jacobi symmetric functions $\mathcal J_{\lambda}$ with $\lambda$ containing the rectangle $(m+1) \times (n+1)$.

\begin{proof} Assume first that  $h$ is not of the form (\ref{1}), (\ref{2}) for all non-negative integers $m,n.$ One can check that under our assumptions on the parameters this implies that 
\begin{equation}
\label{since}V(\lambda, h, -\varepsilon_i) \frac{\mathcal J_{\lambda}(0)}{\mathcal J_{\lambda - \varepsilon_i}(0)}  \neq 0.
\end{equation}

Let $I$ be any $BC$-invariant ideal in $\Lambda,$ which by definition means that $I$ is a linear span of Jacobi symmetric functions $\mathcal J_{\lambda}$
$$
I= Span\{\mathcal J_{\lambda}, \lambda\in \Omega\},
$$
where $\Omega$ is some set of partitions. 

If  $\Omega$ is empty set then by convention $I=0$. Suppose that  $\Omega$ is not empty and $\lambda\in \Omega$. If $\lambda=\emptyset$ then $I=\Lambda$. If $\lambda\ne\emptyset$ then since $p_{1}\mathcal J_{\lambda}\in I$ according to Pieri formula (\ref{pieriinf})  and (\ref{since})
there is  a partition $\mu\subset\lambda$ such that $\mathcal J_{\mu}\in I$ and $\mu$ is obtained from $\lambda$ by removing one box. Therefore by induction in this case we have that $\emptyset\in\Omega$ and $I=\Lambda$.  

Let us now prove that for $h$ of the form (\ref{1}), (\ref{2}) the $BC$-invariant ideal does exist and unique. According to proposition \ref{parameters} (more precisely, the dualities (\ref{change3}) and (\ref{change1})) it is enough to consider only one of those cases.

So we can suppose that  $h= -k(m+2)-(n+1)+\frac12p+q$ for some non-negative integers $m,n$ and consider the set $\Omega$ of all partitions, which contain $(m+1)\times (n+1)$ rectangle $R_{m+1,n+1}.$ Let $I=Span(J_{\lambda}),\lambda \in \Omega$.  We claim that this is an ideal. It is enough to show that this subspace is invariant with respect to multiplication by $E_{r}$ for all $r \in \mathbb Z_{+}.$ From Pieri formula (\ref{pieriinfty}) we see that this will follow if we prove that $V^{-}_{\sigma(L),J,\lambda}(h)V^{+}_{\sigma(J),\lambda}(h)=0$  if $\lambda$ contains $ R_{m+1,n+1}$ and $\lambda+\sigma(J)$ does not. But in that case we have $\lambda_{m+1}=n+1$ and $\sigma_{n+1}=-1$ and therefore by Lemma \ref{infty}  $V^{+}_{\sigma(J),\lambda}(h)=0.$ This proves the existence of the ideal.

 Now let us prove the uniqueness. Let $J$ be an invariant ideal such that $J\ne0,\Lambda$ and $\mathcal J_{\lambda}\in J$. If $\lambda$ does not contain the rectangle $R_{m+1,n+1}$ then we can apply Pieri formula (\ref{pieriinf}) and as before we will get $1\in J$. Therefore $\lambda\supset R_{m+1,n+1}$ and $J\subset I$. Using again the Pieri formula we can get any subdiagram of $\lambda$ containing the rectangle $R_{m+1,n+1},$ which implies that $J=I$.
\end{proof}
\end{thm}

We will describe explicitly the corresponding homomorphism $\Lambda \rightarrow \Lambda/ I$ using the following infinite dimensional analogue of Okounkov's binomial formula.

\begin{proposition} The infinite dimensional Jacobi symmetric functions  can be written in terms of Jack polynomials as follows
\begin{equation}
\label{jinf}
\mathcal J_{\lambda}(u;k,p,q,h)=\sum_{\mu\subseteq\lambda}2^{|\mu|}\frac{\mathcal J_{\lambda}(0)I_{\mu}(\lambda;k,h)}{\mathcal J_{\mu}(0)I_{\mu}(\mu;k,h)}P_{\mu}(u; -k)
\end{equation}
where $P_{\mu}(u; -k)$ is Jack symmetric function corresponding to the partition $\mu$, $ I_{\mu}$ is the interpolation function defined in Proposition \ref{intinf}.
\end{proposition}

The proof follows directly from Theorem \ref{binomial} by the same arguments as in the proof of Theorem \ref{HaCh}.

{\bf Remark.} Using the difference version of Okounkov's formula Rains \cite{Rains} introduced the Koornwinder polynomials of infinitely many variables. The formula (\ref{jinf}) shows that Jacobi symmetric functions are the limiting case of these polynomials.

\begin{proposition}
\label{osob}
When
$$
(h+km+n+\frac12p+q)(h+k(m+1)+n+\frac12p-\frac12)=0
$$
the corresponding ideal is the linear span of Jack polynomials $P_{\lambda}$, where $\lambda$ contains the  rectangle  of the size $(m+1)\times(n+1)$. The corresponding homomorphism is given by the formula
\begin{equation}\label{phi1}
\varphi_{m,n}(p_{a})=\sum_{i=1}^m u_{i}^a+k^{-1}\sum_{j=1}^n v_{j}^{a}
\end{equation}

In the case when 
$$
(h+k(m+2)+(n+1)-\frac12p-q)(h+k(m+1)+n-\frac12p+\frac12)=0
$$
the corresponding ideal is the linear span of the polynomials $\theta(P_{\lambda})$, where $\lambda$ contains the  rectangle  of the size $(m+1)\times(n+1)$. The corresponding homomorphism is given by
\begin{equation}\label{phi2}
\tilde\varphi_{m,n}(p_{a})=\sum_{i=1}^m u_{i}^a+k^{-1}\sum_{j=1}^n v_{j}^{a}+(-2)^a\frac{2q-2k-1}{2k}.
\end{equation}
\end{proposition}
\begin{proof} 

Because of the proposition \ref{parameters} it is enough to investigate only one of the cases. 
Assume that $h= -km-n-\frac12p-q$ for some non-negative $m,n$ and consider the set $\Omega$ of all partitions, which contain $(m+1)\times (n+1)$ rectangle $R_{m+1,n+1}.$ 
It was proven in \cite{SV1} that   $I=Span\{P_{\lambda},\lambda\in\Omega\}$ is the kernel of $\varphi$ defined by (\ref{phi1}) and therefore is an ideal. We claim that it is also $BC$-invariant. 

More precisely, we claim that $I=Span\{\mathcal J_{\lambda},\lambda\in\Omega\}$. To prove this it is enough to show that for such $h$
in the sum at the right hand side of (\ref{jinf}) there are no terms $P_{\mu}(u; -k)$ with $\mu_{m+1}\le n.$
But this follows from the fact that for $h= -km-n-\frac12p-q$ $$\frac{C_{\lambda}^{0}(h+\frac12 p + q)}{C_{\mu}^{0}(h+\frac12 p + q)}=\frac{C_{\lambda}^{0}(-km-n)}{C_{\mu}^{0}(-km-n)} = 0$$ if $\lambda_{m+1} >n$ and $\mu_{m+1}\le n.$ 
\end{proof}

\section{Deformed CMS operator of type $BC(m,n)$ as a restriction}

The main goal of this section is to show that the gauged version of the deformed CMS operator (\ref{bcnm}) related to generalised root system of type $BC(m,n)$ is a restriction of $BC_{\infty}$ operator $\mathcal L^{(k,p,q,h)}.$ This implies quantum integrability of this system and allows us to generalise the main results about the Jacobi polynomials to the deformed case.

Recall that generalised root system $R$ of type $BC(m,n)$ is the union 
\begin{equation}
\label{R}
R=R_{real}\cup  R_{iso}
\end{equation}
of real (non-isotropic) roots
$$
R_{real}=\{\pm\varepsilon_{i} ,\:\pm2\varepsilon_{i},\:\pm\varepsilon_{i}\pm\varepsilon_{j},\, \pm\delta_{p},\, \pm2\delta_{p}, \:\pm\delta_{p}\pm\delta_{q},\: 1\le i\ne j\le m,\; 1\le p\ne q\le n\} 
$$
and isotropic roots
$$
 R_{iso}=\{\pm\varepsilon_{i}\pm\delta_{p}\},
$$
where the bilinear form is defined by
$$
(\varepsilon_{i},\varepsilon_{i})=1,\: 
(\delta_{p},\delta_{p})=-1,\:  (\varepsilon_{i},\delta_{p})=0.
$$
The Weyl group $W_0=\left(S_{m}\ltimes\mathbb{Z}_{2}^m\right)\times\left( S_{n}\ltimes\mathbb{Z}_{2}^n\right)$ acts on the weights by separately  permuting $\varepsilon_{i},\; j=1,\dots,m$ and
$\delta_{p},\; p=1,\dots,n$ and changing their signs.
 
 Consider the deformations of this root system in the sense of \cite{SV}:
 the roots remain the same, but acquire the multiplicities $m({\alpha}),$ 
 what changed is the bilinear form. In the $BC(m,n)$ case the admissible deformations have the form \cite{SV}:
 $$
 (\varepsilon_{i},\varepsilon_{i})=1,\: 
(\delta_{p},\delta_{p})=k,\:  (\varepsilon_{i},\delta_{p})=0.
$$
$$
m(\pm\delta_{p}\pm\varepsilon_{i})=1,\: m(\pm\varepsilon_{i})=p,\: m(\pm2\varepsilon_{i})=q,\: m(\pm\delta_{p})=r,
$$
\begin{equation}
\label{defroot}
m(\pm2\delta_{p})=s,\:m(\pm\varepsilon_{i}\pm\varepsilon_{j})=k,\: m(\pm\delta_{p}\pm\delta_{q})=k^{-1}
\end{equation}
with the following relations to be fulfilled
 $$
p=k r,\: 2q+1=k(2s+1).
 $$
 
The corresponding deformed version of the CMS operator (\ref{CMS}) has the form
$$
\mathcal L^{(m,n)}=\sum_{i=1}^m (x_i\partial_{i})^2+k\sum_{\alpha=1}^n (y_{\alpha}\partial_{\alpha})^2
-k\sum_{i < j}^m \left(\frac{x_{i}+x_{j}}{x_{i}-x_{j}}(x_i\partial_{i}-x_j\partial_{j})+\frac{x_{i}x_{j}+1}{x_{i}x_{j}-1}(x_i\partial_{i}+x_j\partial_{j})\right)
$$
$$
-\sum_{\alpha < \beta}^n \left(\frac{y_{\alpha}+y_{\beta}}{y_{\alpha}-y_{\beta}}(y_{\alpha}\partial_{\alpha}-y_{\beta}\partial_{\beta})+\frac{y_{\alpha}y_{\beta}+1}{y_{\alpha}y_{\beta}-1}(y_{\alpha}\partial_{\alpha}+y_{\beta}\partial_{\beta})\right)
$$
$$
-\sum_{i=1}^m\left(p\frac{x_{i}+1}{x_{i}-1}+2q\frac{x^2_{i}+1}{x^2_{i}-1}\right)x_i\partial_{i}-k\sum_{\alpha=1}^n\left(r\frac{y_{\alpha}+1}{y_{\alpha}-1}+2s\sum_{\alpha=1}^n\frac{y^2_{\alpha}+1}{y^2_{\alpha}-1}\right)y_{\alpha}\partial_{\alpha}
$$
$$
-\sum_{i,\alpha}\left(\frac{x_i+y_{\alpha}}{x_i-y_{\alpha}}(x_i\partial_{i}-ky_{\alpha}\partial_{\alpha})+\frac{x_iy_{\alpha}+1}{x_iy_{\alpha}-1}(x_i\partial_{i}+ky_{\alpha}\partial_{\alpha})\right),
$$
where $\partial_{i}=\frac{\partial}{\partial x_{i}},\:\partial_{\alpha}=\frac{\partial}{\partial y_{\alpha}}.$

Introduce the new coordinates $$u_{i}=\frac12(x_{i}+x^{-1}_{i}-2),\:v_{\alpha}=\frac12(y_{\alpha}+y^{-1}_{\alpha}-2).$$ In these coordinates the operator $\mathcal{L}^{(m,n)}$ has the following expression (cf. (\ref{CMS1})):

$$
\mathcal L^{(m,n)}=\sum_{i=1}^m (u_i\partial_{i})^2+k\sum_{\alpha=1}^n (v_{\alpha}\partial_{\alpha})^2
-k\sum_{i < j}^m \frac{u_{i}+u_{j}}{u_{i}-u_{j}}(u_i\partial_{i}-u_j\partial_{j})$$
$$
-\sum_{\alpha< \beta}^n \frac{v_{\alpha}+v_{\beta}}{v_{\alpha}-v_{\beta}}(v_{\alpha}\partial_{\alpha}-v_{\beta}\partial_{\beta})
-\sum_{i,\alpha}\frac{u_{i}+v_{\alpha}}{u_{i}-v_{\alpha}}(u_i\partial_{i}-kv_{\alpha}\partial_{\alpha})
$$
$$
-(n+k(m-1)+p+2q)\left(\sum_{i=1}^m u_i\partial_{i}+\sum_{\alpha=1}^n v_{\alpha}\partial_{\alpha}\right)-(1+2p+2q)\left(\sum_{i=1}^m\partial_{i}+\sum_{r=1}^n \partial_{\alpha}\right)
$$
$$
+2\sum_{i=1}^m \partial_{i} (u_i \partial_{i})+2k\sum_{\alpha=1}^n \partial_{\alpha} (v_{\alpha}\partial_{\alpha})-4k\sum_{i<j}^m \frac{u_i\partial_{i}-u_j\partial_{j}}{u_{i}-u_{j}}
$$
\begin{equation}
\label{defCMS}
-4\sum_{\alpha< \beta}^n\frac{v_{\alpha}\partial_{\alpha}-v_{\beta}\partial_{\beta}}{v_{\alpha}-v_{\beta}}
-4\sum_{i,\alpha}\frac{u_i\partial_{i}-kv_{\alpha}\partial_{\alpha}}{u_{i}-v_{\alpha}}.
\end{equation}

 Let $P_{m,n}=\mathbb C[u_{1},\dots,u_{m},v_{1},\dots,v_{n}]$ be the polynomial algebra in $n+m$ independent  variables. Denote by $\Lambda_{m,n,k}$ its subalgebra, consisting of polynomials, which are symmetric in $u_{1},\dots,u_{m}$ and $v_{1},\dots, v_{n}$ separately  and satisfy the conditions
$$
u_{i}\frac{\partial f}{\partial u_{i}}-kv_{\alpha}\frac{\partial f}{\partial u_{\alpha}}=0
$$
on the hyperplanes  $u_{i}=v_{\alpha}$ for all $i=1,\dots,m$ and $\alpha=1,\dots, n.$ For generic values of parameter $k$ (more precisely, if $k$ is not a positive rational number) this algebra is generated by the {\it deformed power sums} 
$$p_a(u,v,k)= \sum_{i=1}^m u_{i}^a+k^{-1}\sum_{\alpha=1}^{n}v_{\alpha}^a,\,\, \:a=1,2,\dots$$
(see Theorem 2 in \cite{SV}). 
Consider a homomorphism $\varphi_{m,n}: \Lambda\longrightarrow\Lambda_{m,n,k}$ 
defined by $$\varphi_{m,n} (p_{a})= p_a(u,v,k).$$

The main result of this section is the following theorem (cf. Theorem 1 in \cite{SV1}).

\begin{thm}  
\label{restric}
The following diagram is commutative for $h=-km-n-\frac{1}{2}p-q$ and generic values of parameter $k$:
\begin{equation} \label{commdia}
\begin{array}{ccc}
\Lambda&\stackrel{{\mathcal L^{(k,p,q,h)}}_{}}{\longrightarrow}&\Lambda
\\ \downarrow \lefteqn{\varphi_{m,n}}& &\downarrow \lefteqn{\varphi_{m,n}}\\
\Lambda_{m,n,k}&\stackrel{{\mathcal
L}^{(m,n)}}{\longrightarrow}&\Lambda_{m,n,k} \\
\end{array}
\end{equation}
In other words, the deformed CMS operator (\ref{defCMS}) is a restriction of the $BC_{\infty}$ CMS operator onto the corresponding subvariety $\mathcal D(m,n,k) = Spec \, \Lambda_{m,n,k}.$
\begin{proof} 
The proof is similar to the proof of Theorem (\ref{main}). Let us split the operator $\mathcal
L^{(m,n)}$  as the sum:
$$
\mathcal L^{(m,n)}=\mathcal A^{(m,n)}+\mathcal B^{(m,n)},
$$
where
$$
\mathcal A^{(m,n)}=\sum_{i=1}^m (u_i\partial_{i})^2+k\sum_{\alpha=1}^n (v_{\alpha}\partial_{\alpha})^2
-k\sum_{i < j}^m \frac{u_{i}+u_{j}}{u_{i}-u_{j}}(u_i\partial_{i}-u_j\partial_{j})$$
$$
-\sum_{\alpha< \beta}^n \frac{v_{\alpha}+v_{\beta}}{v_{\alpha}-v_{\beta}}(v_{\alpha}\partial_{\alpha}-v_{\beta}\partial_{\beta})
-\sum_{i,\alpha}\frac{u_{i}+v_{\alpha}}{u_{i}-v_{\alpha}}(u_i\partial_{i}-kv_{\alpha}\partial_{\alpha})
$$
$$
-(n+k(m-1)+p+2q)\left(\sum_{i=1}^m u_i\partial_{i}+\sum_{\alpha=1}^n v_{\alpha}\partial_{\alpha}\right),
$$
$$
\mathcal B^{(m,n)}=
2\sum_{i=1}^m \partial_{i} (u_i \partial_{i})+2k\sum_{\alpha=1}^n \partial_{\alpha} (v_{\alpha}\partial_{\alpha})-4k\sum_{i<j}^m \frac{u_i\partial_{i}-u_j\partial_{j}}{u_{i}-u_{j}}
$$
$$
-4\sum_{\alpha< \beta}^n\frac{v_{\alpha}\partial_{\alpha}-v_{\beta}\partial_{\beta}}{v_{\alpha}-v_{\beta}}
-4\sum_{i,\alpha}\frac{u_i\partial_{i}-kv_{\alpha}\partial_{\alpha}}{u_{i}-v_{\alpha}}
$$
$$
-(1+2p+2q)\left(\sum_{i=1}^m\partial_{i}+\sum_{r=1}^n \partial_{\alpha}\right).
$$
The claim follows from the following formulas:
 $$
\mathcal A^{(m,n)} (\hat p_{a})=(1+k)a^2\hat p_{a}-ka\sum_{r=0}^{a-1}\hat p_{r}\hat p_{a-r}-a(n+k(m-1)+p+2q)\hat p_{a},
 $$
$$
\mathcal A^{(m,n)}(\hat p_{a}\hat p_{b})=\mathcal A^{(m,n)}(\hat p_{a})\hat p_{b}+\hat p_{a}\mathcal A^{(m,n)}(\hat p_{b})+2ab\hat p_{a+b},
 $$
  $$
  \mathcal B^{(m,n)}(\hat p_{a})=2(1+k)a^2\hat p_{a-1}-2ka\sum_{r=0}^{a-1}\hat p_{r}\hat p_{a-r-1}-a(1+2p+2q)\hat p_{a-1},
 $$
$$
 \mathcal B^{(m,n)}(\hat p_{a}\hat p_{b})=  \mathcal B^{(m,n)}(\hat p_{a})\hat p_{b}+\hat p_{a} \mathcal B^{(m,n)}(\hat p_{b})+4ab\hat p_{a+b-1}.
 $$
where $\hat p_a= p_a(u,v,k)$ and  $\hat p_0=m+k^{-1}n.$
They can be checked directly with the use
the following additional identities:
$$
\sum_{i,\alpha}(u_{i}+v_{\alpha})\frac{u_{i}^a-v_{\alpha}^a}{u_{i}-v_{\alpha}}=\sum_{r=0}^{a-1}\left(p_{r}(u)p_{a-r}(v)+p_{r}(v)p_{a-r}(u)\right),
$$
$$
\sum_{i,\alpha}\frac{u_{i}^a-v_{\alpha}^a}{u_{i}-v_{\alpha}}=\sum_{r=0}^{a-1}p_{r}(u)p_{a-r-1}(v),
$$
where $p_0(u)=m, \, p_0(v)=n.$ 

Note that from these formulas it follows that for generic $k$ the operator $\mathcal
L^{m,n}$ maps the algebra $\Lambda_{m,n,k}$ into itself.  The rest of the proof is the same as in Theorem \ref{main}.
\end{proof}
\end{thm}

{\bf Remark 1.} One can show that the diagram in the theorem is commutative and the operator $\mathcal
L^{m,n}$ preserves the algebra $\Lambda_{m,n,k}$ for {\bf all} values of parameter $k$ (see \cite{SV1}, where this is proved in the  $A(m,n)$ case).

{\bf Remark 2.} Because of the symmetry (\ref{change2})
the theorem remains true when $h=-k(m+1)-n - \frac{1}{2}p+\frac{1}{2}$ if we change the parameters in the operator $\mathcal
L^{m,n}$ according to the formula (\ref{sym1}). The situation for other two exceptional values (\ref{2}) of parameter $h$ from the theorem \ref{singular} still needs to be understood. We believe that the restriction is again the deformed $BC(m,n)$ operator written in a different coordinate system.

As a corollary we have quantum integrability of the deformed $BC(m,n)$ system. Let the algebra $\Lambda^{(k,h)}$ and the operators $\mathcal L^{(k,p,q,h)}$ be the same as in Theorem \ref{HaCh} from section 3. 

\begin{corollary}
For $h=-km-n-\frac12 p -q$
and any $f\in \Lambda^{(k,h)}$ there exists a unique differential operator $\mathcal L^{(m,n)}_{f}$ on $\Lambda_{m,n,k}$ such that the following diagram is commutative
 $$
\begin{array}{ccc}
\Lambda&\stackrel{\mathcal L^{(k,p,q,h)}_{f}}{\longrightarrow}& \Lambda\\
\downarrow \lefteqn{\varphi_{m,n}}& &\downarrow \lefteqn{\varphi_{m,n}}\\
\Lambda_{m,n,k}&\stackrel{\mathcal L^{(m,n)}_{f}}{\longrightarrow}&\Lambda_{m,n,k}.
 \\
\end{array}
$$ 
The operators $\mathcal L^{(m,n)}_{f}$ commute with each other.
\end{corollary}

In the next section we discuss the corresponding algebra of quantum integrals in more detail.

\section{Algebra of quantum integrals of the deformed $BC(m,n)$ system}

Consider the deformed $BC(m,n)$ root system (\ref{defroot}) and 
choose the following system of positive roots:
$$
R^{+}_{real}= \{\varepsilon_{i} ,\:2\varepsilon_{i} ,\:\varepsilon_{i}\pm\varepsilon_{j},\delta_{p} ,\:2\delta_{p} ,\:\delta_{p}\pm\delta_{q},\: 1\le i< j\le m,\; 1\le p< q\le n\},
$$
\begin{equation}
\label{roots}
R^{+}_{iso}=\{\delta_{p}\pm\varepsilon_{i}\}.
\end{equation}
Let $\rho$ be the corresponding deformed version of the half-sum of positive roots:
$$
 \rho=\frac12\sum_{\alpha\in R^{+}}m(\alpha)\alpha,
 $$
which in our case has the form 
$$
\rho=\sum_{i=1}^m\left(\frac12p+q+k(m-i)\right)\varepsilon_{i}+\sum_{j=1}^n\left(\frac12r+s+m+k^{-1}(n-j)\right)\delta_{j}
$$

\begin{equation}
\label{rhodef}
=-\sum_{i=1}^m (h+n+ki)\varepsilon_i - k^{-1}\sum_{j=1}^n (h+\frac{1}{2}k-\frac{1}{2}+j)\delta_{j},
\end{equation}
where $h=-km-n-\frac{1}{2}p-q.$

Consider now the following deformed analogue of algebra $\Lambda_{N}^{(k,h)}$ from section 3.
Let $\Lambda_{m,n}^{(k,h)} \subset P_{m,n} = \mathbb C[w_1,\dots,w_m, z_1, \dots, z_n]$ be the algebra of polynomials in $\, w_{1},\dots, w_{m}, z_{1},\dots,z_{n},$ which are symmetric separately in variables $(w_{i}+h+n+ki)^2,\, i=1,\dots,m$ and  $(z_{j}+hk^{-1}+\frac{1}{2}-\frac{1}{2}k^{-1}+jk^{-1})^2, \,j=1,\dots,n$ and have the following property:
$$
f(w-\varepsilon_i, z+\delta_j)=f(w, z)
$$
on each hyperplane $w_{i}+k(i-1)=kz_{j}+j-n$ for all $i=1,\dots, m$ and $j=1, \dots, n.$
In terms of the deformed root system $BC(m,n)$ the last condition can be rewritten as
$f(X+\alpha)=f(X),$ when $X=(w,z)$ belongs to the hyperplane $$(X-\rho,\alpha)+\frac{1}{2} (\alpha,\alpha)=0$$
where  $\alpha=\delta_j - \varepsilon_i$ is an isotropic root. The algebra $\Lambda_{m,n}^{(k,h)}$  is isomorphic (by the shift by $\rho$) to the algebra $\Lambda_{R,B}$ introduced in \cite{SV}, when $R$ is the generalised root system of type $BC(m,n).$

Let us denote by $H_{m,n}$ the set of partitions $\lambda$  such that $\lambda_{m+1}\le n.$ This means that the corresponding Young diagrams are contained in the fat $(m,n)$-hook (see \cite{SV}). For such a diagram one can consider two sub-diagrams $\mu$ and $\nu$, where $\nu$ consists of first $n$ columns of $\lambda$ and $\mu$ is the remaining part.
Consider the following embedding $\chi: H_{m,n} \rightarrow \mathbb C^{n+m},$ using the formulas 
\begin{equation}
\label{embed}
\chi (\lambda) = (w(\lambda), z(\lambda)),
\end{equation}
where
\begin{equation}
\label{embedding}
w_i(\lambda)=\mu_{i},\: i=1,\dots,m, \,\, z_j(\lambda)=\nu^{\prime}_{j},\: j=1,\dots,n.
\end{equation}
In other words $w_i$ is the length of $i$-th row of $\mu$ and $z_j$ is the length of $j$-th column of $\lambda \in H_{m,n}.$

 It is easy to see that the image of $\chi$  is dense with respect to Zarisski topology. Therefore considering any element from $\Lambda^{(k,h)}$ as the function on  $H_{m,n}$  we have well defined map 
$$
res_{m,n}: \Lambda^{(k,h)}\longrightarrow P_{m,n}.$$

\begin{thm} 
For generic values of parameter $k$  the image of the restriction homomorphism $res_{m,n}$ coincides with the algebra  $\Lambda_{m,n}^{(k,h)}.$

  \begin{proof} Let us  consider  the following functions
    $$
 f_{l}(w)= \sum_{i\ge 1} (B_{2l}(w_{i}+h+ki+1/2)
 -B_{2l}(h+ki+1/2))
  $$
where $B_{2l}(x)$ are the even Bernoulli polynomials.  These polynomials have the following properties (see e.g. \cite{WW}):
\begin{equation}\label{Berodd}
B_{2l}(x)=B_{2l}(1-x)
\end{equation}
and 
\begin{equation}\label{Bereven}
B_{2l}(x+t)=2l\sum_{i=1}^t (x+i-1)^{2l-1}+B_{2l}(x)
\end{equation}
for any positive integer $t$.

It is easy to check that the functions $f_l(w)$ belong to $\Lambda^{(k,h)}$ and  generate this algebra (cf.  \cite{SV1}, section 3).  When $w=\lambda$ is a partition we have
$$
 f_{l}(\lambda)=2l\sum_{(ij)\in\lambda}(j-1+k(i-1)+h+k+\frac{1}{2})^{2l-1}.
$$

  On the other hand, consider the functions 
  $$
  f^{m,n}_{l}(w,z)=\sum_{i=1}^m\left(B_{2l}(w_{i}+h+n+ki+\frac{1}{2})
  - B_{2l}(h+n+ki+\frac{1}{2})\right)
  $$
  $$
 +k^{2l-1}\sum_{j=1}^n(B_{2l}(z_{j}+k^{-1}h+k^{-1}j-\frac{1}{2}k^{-1}+1)- B_{2l}(k^{-1}h+k^{-1}j-\frac{1}{2}k^{-1}+1)).$$
It is easy to verify that these functions belong to $\Lambda_{m,n}^{(k,h)}$ and generate it for generic values of $k.$ If $w_{i}=\mu_{i}, \,\, z_{j}=\nu^{\prime}_{j}$ are determined as above by some partition $\lambda$ we have
 $$
 f^{m,n}_{l}(\lambda) 
   =2l\sum_{(i,j)\in\mu}(j+n-1+k(i-1)+h+k+\frac{1}{2})^{2l-1}
 $$
   $$
 + 2lk^{2l-1}\sum_{(i,j)\in\nu}(i-1+k^{-1}(j-1)+k^{-1}h+\frac{1}{2}k^{-1}+1)^{2l-1},
  $$
which can be rewritten as  
  $$
 f^{m,n}_{l}(\lambda)=2l\sum_{(i,j)\in\lambda}(j-1+k(i-1)+h+k+\frac{1}{2})^{2l-1}
  $$
and coincides with $f_{l}(\lambda).$ Thus we have shown that $$res_{m,n}  (f_{l}) =  f^{m,n}_{l},$$
which implies the theorem.
\end{proof}
\end{thm}

As a corollary we have the following deformed version of the (inverse) Harish-Chandra homomorphism.
Let $\mathcal D_{m,n}$ be the algebra of quantum integrals of the deformed $BC(m,n)$ system with the Hamiltonian $\mathcal L^{m,n}.$

\begin{corollary} For $h=-km-n-\frac{1}{2}p-q$ and generic values of $k$ there exists a monomorphism $$\chi: \Lambda_{m,n}^{(k,h)} \rightarrow \mathcal D_{m,n}.$$ 
\end{corollary}

An alternative version of this monomorphism was established in \cite{SV} by a complicated recurrent construction of the quantum integrals. The present form gives a more conceptual way to do this.

Now we are going to describe an important linear basis in algebra $\Lambda_{m,n}^{(k,h)},$ 
consisting of the {\it deformed interpolation $BC$ polynomials} $I_{\lambda}(z,w,k,h).$ They can be defined as follows (cf. Proposition \ref{intfinite}).

\begin{proposition} For any partition $\lambda$ such that $\lambda_{m+1}\le n$ there exists a polynomial $I_{\lambda}(z_{1},\dots, z_{n}, \, w_{1},\dots, w_{m}, k, h) \in \Lambda^{(k,h)}_{n,m}$  of degree $deg( I_{\lambda})=2|\lambda|,$ which satisfies the following conditions:
$$I_{\lambda}(z(\mu), w(\mu),k,h)=0,$$  where $\mu \in H_{m,n}$ is any partition such that 
$\lambda\nsubseteq\mu$, $z(\mu)$ and $w(\mu)$ are defined by (\ref{embed}), and normalisation
$$
I_{\lambda}(z(\lambda), w(\lambda),k,h)=\prod_{(i,j)\in\lambda}(1+\lambda_{i}-j-k(\lambda^{\prime}_{j}-i))(2h-1+\lambda_{i}+j+k(\lambda^{\prime}_{j}+i)).
$$
\end{proposition}

The proof easily follows from Proposition \ref{intinf}.

One can give the following deformed analogue of Okounkov's combinatorial formula for interpolation $BC$ polynomials \cite{Ok}.

Let  $T$ be a reverse bitableau of type $(n,m)$ and shape $\lambda$. We  can view $T$ as a filling of a
Young diagram  $\lambda$ by symbols $1<2<\dots
<n<1^{\prime}<2^{\prime}<\dots< m^{\prime}$ with entries
 decreasing weakly downwards in each column and rightwards in each row; additionally 
entries $1, 2\dots , n$ decrease strictly  downwards in each column and
entries $1^{\prime},2^{\prime}\dots, m^{\prime}$ decrease strictly
rightwards in each row. Let $T_{1}$ be a subtableau  in $T$ containing
all symbols $1^{\prime},2^{\prime}\dots, m^{\prime}$ and
$T_{0}=T-T_{1}$ and $\mu$ is the shape of $T_{1}$.

Introduce the polynomials
\begin{equation}
\label{ihat}
\hat I_{\lambda}(z_1, \dots, z_n, \, w_1, \dots, w_m, k, h) = \sum_{T} \varphi_{T}(-k) \prod_{\Box \in \lambda} f_T(\Box),
\end{equation}
where 
$$f_T(\Box) = (z_{T(\Box)}+h + kT(\Box))^2 -(c_{\lambda}^0(\Box, 0) + h + kT(\Box))^2$$
if $T(\Box)=1,\dots n$ and
$$f_T(\Box) = k^2[(w_{|T(\Box)|}+\hat h +n + k^{-1}|T(\Box)|)^2 -(k^{-1}c_{\lambda}^0(\Box, 0) + \hat h + n+ k^{-1}|T(\Box)|)^2$$
if $T(\Box)=1^{\prime},\dots m^{\prime}$ in which case we denote $|T(\Box)|=1,\dots m$ respectively.
Let $\varphi_{T}(-k)$ be the same weight as for super Jack polynomials (see formula (41) in \cite{SV1}).

\begin{proposition} The deformed interpolation $BC$ polynomials can be given by the following  formula:
$$
I_{\lambda}(z_1, \dots, z_n, w_{1},\dots, w_{m}, k,h) = d_{\lambda} \hat I_{\lambda^{\prime}}(z_1, \dots, z_n, w_{1},\dots, w_{m}, k^{-1}, \hat h)$$
 where $\hat I_{\lambda}(z, \, w, k, h)$ is given above by combinatorial formula (\ref{ihat}) and $$d_{\lambda} = (-1)^{|\lambda|} k^{2|\lambda|} \frac{C_{\lambda}^{-}(1)}{C_{\lambda}^{-}(-k)}.$$ \end{proposition}

The proof is based on the Okounkov's results \cite{Ok} and is similar to the super-Jack case (see Theorem 12 in \cite{SV1}). The appearance of the dual parameters and transposed Young diagram is a consequence of our choice of coordinates.

\section{Super Jacobi polynomials}

In this section we introduce and investigate the deformed version of the Jacobi polynomials.
In particular, we prove the corresponding analogues of Pieri, Opdam's and Okounkov's formulae.

We define the {\it super Jacobi polynomials} $\mathcal J_{\lambda}(u,v; k,p,q)$ as the image of 
the Jacobi symmetric functions $\mathcal J_{\lambda}(z; k,p,q,h)$ under the homomorphism (\ref{phi1}):
\begin{equation}
\label{supjac}
\mathcal J_{\lambda}(u,v; k,p,q)=\varphi_{m,n}(\mathcal J_{\lambda}(z; k,p,q,h)),
\end{equation}
where $h=-km-n-\frac{1}{2}p - q.$ 

From Theorem \ref{restric} and Proposition \ref{osob} it follows that $\mathcal J_{\lambda}(u,v; k,p,q)$ are the eigenfunctions of the deformed operator $\mathcal L^{m,n}$ and form a basis of $\Lambda_{m,n}$ if we consider all the partitions $\lambda \in H_{m,n}$ from the fat $(m,n)$-hook (i.e. with
$\lambda_{m+1}\le n$).

It is convenient also to use the normalised version of super Jacobi polynomials. To define them we introduce the following partial order on the pairs of partitions: $(\mu,\nu) \preceq (\tilde \mu, \tilde \nu)$ if the corresponding
$X=(x_1,\dots,x_{m+n}) = (\nu,\mu)$ with $x_1=\nu_1,\dots, x_n=\nu_n, x_{n+1}=\mu_1, \dots, x_{m+n}=\mu_m$ and $\tilde X=(\tilde \nu, \tilde \mu)$ satisfy the inequalities
\begin{equation}
\label{order}
x_1+\dots+x_i \leq \tilde x_1 + \dots + \tilde x_i
\end{equation}
for all $i=1, \dots, m+n.$ Note that in these formulas $\nu$ goes first.
One can check that the highest term in the super Jacobi polynomial $\mathcal J_{\lambda}(u,v; k,p,q)$ with respect to this order is $m_{\mu}(u)m_{\nu}(v)$ with the coefficient given explicitly by the following

\begin{proposition}\label{norm}
The highest term in super Jacobi polynomial $\mathcal J_{\lambda}(u,v; k,p,q)$ has the form $a_{\lambda}m_{\mu}(u)m_{\nu}(v)$ with
\begin{equation}
\label{al}
a_{\lambda}=(-1)^{|\nu|}2^{|\lambda|} \frac{C^{-}_{\lambda}(1) C^{-}_{\mu}(-k) }{C^{-}_{\lambda}(-k) C^{-}_{\mu}(1)},
\end{equation}
where as before the sub-diagram $\nu \subset \lambda$ consists of the first $n$ columns and $\mu=\lambda\setminus\nu.$
\end{proposition}

The proof follows from a similar result for super Jack polynomials. Recall that {\it super-Jack polynomial} $SP_{\lambda}(u,v; -k)$ is defined as the image of the Jack symmetric function $P_{\lambda}(x; -k)$ under the homomorphism $\varphi_{m,n}$ (see \cite{SV1}).

\begin{lemma}\label{norm}
The highest term of the super Jack polynomial $SP_{\lambda}(u,v; -k)$ has the form  $b_{\lambda} m_{\mu}(u)m_{\nu}(v)$ with $b_{\lambda}= a_{\lambda} 2^{-|\lambda|}$ and $a_{\lambda}$ given above by (\ref{al}). 
\end{lemma}

This follows directly from the definition of the function $C^{-}_{\lambda}:$
$$C^{-}_{\lambda}(x)=\prod_{(ij)\in\lambda}\left(\lambda_{i}-j-k(\lambda^{\prime}_{j}-i)+x\right),$$
 the formulas (7.11'), (7.13'), (10.10), (10.11) from Macdonald \cite{Ma} and formula 
(25) from \cite{SV1}.

Now we define the 
{\it normalised super Jack polynomials} as
$$
SP^*_{\lambda}(u,v; -k)=b_{\lambda}^{-1}SP_{\lambda}(u,v; -k).
$$
and the
{\it normalised super Jacobi polynomials} as
$$
 \mathcal J^*_{\lambda}(u,v; k,p,q)=a_{\lambda}^{-1}\mathcal J_{\lambda}(u,v; k,p,q).
$$

We have the following generalisation of Okounkov's formula to the deformed case.

\begin{thm} The normalised super Jacobi polynomials can be written in terms of super Jack polynomials as follows:
\begin{equation}
\label{sbinom}
{\mathcal J}^*_{\lambda}(u,v; k,p,q)=\sum_{\mu\subseteq\lambda} \frac{{\mathcal{J}^*_{\lambda}}(0)I_{\mu}(w(\lambda),z(\lambda); k,h)}{{\mathcal {J}}^*_{\mu}(0)I_{\mu}(w(\mu),z(\mu);k,h)}SP^*_{\mu}(u,v; -k),
\end{equation}
where $\lambda \in H_{m,n},\,\,h=-km-n-\frac12p-q$ and ${\mathcal {J}}^*_{\mu}(0)={\mathcal {J}}^*_{\mu}(0,0;k,p,q).$
\end{thm}

The proof follows from the fact that the super Jack and corresponding super Jacobi polynomials have the same highest term with respect to the partial order introduced above.

We are going to present now an explicit expression for the values ${\mathcal {J}}^*_{\mu}(0).$
Let
\begin{equation}\label{chi}
\chi({\lambda})=z_{1}(\lambda)\delta_{1}+\dots+z_{n}(\lambda)\delta_{n}+w_{1}(\lambda)\varepsilon_{1}+\dots+w_{m}(\lambda)\varepsilon_{m},
\end{equation}
where $z_i(\lambda),\, w_j(\lambda)$ are defined above by (\ref{embed}),
and $\rho$ be the deformed version of half-sum of positive roots given by (\ref{rhodef}).
Introduce \begin{equation}\label{sUpsilon}
\Upsilon(\lambda, \alpha, k)=\frac{\Gamma((\chi({\lambda})-\rho,\alpha^{\vee})-m_{\alpha}-\frac12m_{\alpha/2})\Gamma((-\rho,\alpha^{\vee})-\frac12m_{\alpha/2})}{\Gamma((\chi({\lambda})-\rho,\alpha^{\vee})-\frac12m_{\alpha/2})\Gamma((-\rho,\alpha^{\vee})-m_{\alpha}-\frac12m_{\alpha/2})},
\end{equation}
where the deformed inner product  $( , )$ and the multiplicities are given by (\ref{defroot}) and $\alpha^{\vee}=\frac{2\alpha}{(\alpha,\alpha)} $ (cf. formula (\ref{Upsilon}) above).

We have the following generalisation of Opdam's formula (\ref{j0}) to the super case.
\begin {thm} 
The value of the normalised super Jacobi polynomials at $0$ is given by
$$
\mathcal{J}^*_{\lambda}(0,0;k,p,q)=\prod_{\alpha\in R^{+}}\Upsilon(\lambda, \alpha, k),
$$
where the product is taken over the positive roots (\ref{roots}).
\begin{proof}
We split the positive roots into three groups: 
$$
R^{+}_{\delta}= \{\delta_{p} ,\:2\delta_{p} ,\:\delta_{p}\pm\delta_{q},\:  1\le p< q\le n\},
$$
$$
R^{+}_{\epsilon}= \{\varepsilon_{i} ,\:2\varepsilon_{i} ,\:\varepsilon_{i}\pm\varepsilon_{j},\: 1\le i< j\le m\},
$$
$$
R^{+}_{iso}=\{\delta_{p}\pm\varepsilon_{i},\:\delta_{p}\},
$$
so that
$$
\prod_{\alpha\in R^{+}}\Upsilon(\lambda, \alpha, k)=\prod_{\alpha\in R^{+}_{\delta}}\Upsilon(\lambda, \alpha, k)\prod_{\alpha\in R^{+}_{\varepsilon}}\Upsilon(\lambda, \alpha, k)\prod_{\alpha\in R^{+}_{iso}}\Upsilon(\lambda, \alpha, k).
$$
Using Proposition \ref{combin}
we have 
$$
\prod_{\alpha\in R^{+}_{\varepsilon}}\Upsilon(\lambda, \alpha, k)=4^{|\mu|}\frac{C_{\mu}^{0}(-km)C_{\mu}^{0}(x+n)}{C_{\mu}^{-}(-k)C_{\mu}^{+}(y+n-1)},
$$
$$
\prod_{\alpha\in R^{+}_{\delta}}\Upsilon(\lambda, \alpha, k)=(-4)^{|\nu|}\frac{C_{\nu}^{0}(-n)C_{\nu}^{0}(x)C_{\nu}^0(y+k)}{C_{\nu}^{-}(1)C_{\nu}^{+}(y-n-1)C_{\nu}^0(y+k+km)},
$$
where
$$
x=k(1-m)-n-p-q+\frac12,\quad y=-n-2km-p-2q.
$$
We have also by definition
$$
\prod_{\alpha\in R^{+}_{iso}}\Upsilon(\lambda, \alpha, k)=\prod_{i=1}^m\prod_{j=1}^n\frac{j-ki-n-1}{k\nu^{\prime}_{j}-\mu_{i}+j-ki-n-1}\frac{j+ki+y-1}{k\nu^{\prime}_{j}+\mu_{i}+j+ki+y-1}.
$$
From proposition \ref{jzero} we have 
$$
\mathcal{J}^*_{\lambda}(0,0;k,p,q) =4^{|\lambda|}a_{\lambda}^{-1}\frac{C_{\lambda}^{0}(-km-n)C_{\lambda}^{0}(x)}{C_{\lambda}^{-}(-k)C_{\lambda}^{+}(y-n-1)}.
$$
The theorem follows from the following two identities:
$$
\frac{C_{\lambda}^{0}(-km-n)}{a_{\lambda}C_{\lambda}^{-}(-k)}=\frac{C_{\mu}^{0}(-km)C_{\nu}^{0}(-n)}{C_{\mu}^{-}(-k)C_{\nu}^{-}(1)}\prod_{i=1}^m\prod_{j=1}^n\frac{j-ki-n-1}{k\nu^{\prime}_{j}-\mu_{i}+j-ki-n-1}
$$
and
$$
\frac{C_{\lambda}^{0}(x)}{C_{\lambda}^{+}(y-n-1)}=\frac{C_{\mu}^{0}(x+n)}{C_{\mu}^{+}(y+n-1)}
\frac{C_{\nu}^{0}(x)C_{\nu}^{0}(y+k)}{C_{\nu}^{+}(y-n-1)C_{\nu}^0(y+k+km)}
$$
$$
\prod_{i=1}^m\prod_{j=1}^n\frac{j+ki+y-1}{k\nu^{\prime}_{j}+\mu_{i}+j+ki+y-1}.
$$
Let us prove the first identity. It can be rewritten as 
$$
\frac{C_{\lambda}^{0}(-km-n)C_{\mu}^{-}(1)C_{\nu}^{-}(1)}{C_{\mu}^{0}(-km)C_{\nu}^{0}(-n) C_{\lambda}^{-}(1)}=\prod_{i=1}^m\prod_{j=1}^n\frac{j-ki-n-1}{k\nu^{\prime}_{j}-\mu_{i}+j-ki-n-1}.
$$
To prove this we note that
$$
\prod_{i=1}^l\frac{z+k(i-m-1)}{z+k(i-1)}\equiv\prod_{j=1}^m\frac{z-kj}{z+k(l-j)}
$$
for any two  nonnegative integers  $l,m.$ Using this one can show that
$$
\frac{C_{\lambda}^{0}(-km-n)}{C_{\mu}^{0}(-km)C_{\nu}^{0}(-n)}=\prod_{i=1}^m\prod_{j=1}^n\frac{j-ki-n-1}{k\nu^{\prime}_{j}+j-ki-n-1}.
$$
We can also check directly that
$$
\frac{C_{\lambda}^{-}(1)}{C_{\mu}^{-}(1)C_{\nu}^{-}(1)}=\prod_{i=1}^m\prod_{j=1}^n\frac{k\nu^{\prime}_{j}-\mu_{i}+j-ki-n-1}{k\nu^{\prime}_{j}+j-ki-n-1},
$$
which implies the first identity.

Now let us prove the second identity. It is easy to verify that it is equivalent to the following equality
$$
\frac{C_{\mu}^{+}(y+n-1)C_{\nu}^{+}(y-n-1)C_{\nu}^0(y+k+km)}{C_{\lambda}^{+}(y-n-1)C_{\nu}^{0}(y+k)}=\prod_{i=1}^m\prod_{j=1}^n\frac{j+ki+y-1}{k\nu^{\prime}_{j}+\mu_{i}+j+ki+y-1}.
$$
Using the identity
$$
\prod_{i=1}^l\frac{z+k(i-1)}{z+k(i-1+m)}=\prod_{j=1}^m\frac{z+k(j-1)}{z+k(l+j-1)}
$$
for arbitrary  nonnegative integers $l,m$ one can show that
$$
\frac{C_{\nu}^0(y+k(1+m))}{C_{\nu}^0(y+k)}=\prod_{i=1}^m\prod_{j=1}^n\frac{j+ki+y-1}{k\nu^{\prime}_{j}+j+ki+y-1}.
$$
A direct check gives also
$$
\frac{C_{\mu}^{+}(y+n-1)C_{\nu}^{+}(y-n-1)}{C_{\lambda}^{+}(y-n-1)}=\prod_{i=1}^m\prod_{j=1}^n\frac{k\nu^{\prime}_{j}+j+ki+y-1}{k\nu^{\prime}_{j}+j+\mu_{i}+ki+y-1}.
$$
Theorem is proved.
\end{proof}
\end{thm}

Now we are going to present a deformed version of the Pieri formula.
When $n=1$ a related result  was found by M. Feigin \cite{F2}, who proved the bispectrality of the deformed CMS problem (\ref{bcnm})  in that case.

Consider the set $$\mathcal O=\{\pm \delta_i,\, \pm \varepsilon_j|\,\,i=1,\dots,n,\, j=1,\dots,m\}.$$
For a given partition $\lambda$ we say that $\beta \in \mathcal O$ is admissible if $\chi(\lambda)+\beta = \chi(\tilde\lambda)$ for some partition  $\tilde \lambda \in H_{m,n}.$ We denote the corresponding partition $\tilde \lambda$ as $\lambda + \beta.$

Let $R$ be the $BC(m,n)$ root system (\ref{R})
and $<,>$ denote the standard inner product: 
$<\varepsilon_i, \varepsilon_j>=<\delta_i,\delta_j>=0$ when $i\neq j$, $<\varepsilon_i, \varepsilon_i>=<\delta_j,\delta_j>=1$ and $<\varepsilon_i, \delta_j>=0.$
Introduce the function
 \begin{equation}
\label{SV}
V(\lambda,\beta)=(\beta,\beta)^{-1}\prod_{\alpha\in R,<\alpha,\beta> > 0}\frac{\Upsilon(\lambda+\beta,\alpha,k)}{\Upsilon(\lambda,\alpha,k)},
 \end{equation}
 where $\Upsilon(\lambda,\alpha,k)$ is defined by formula (\ref{sUpsilon}).
It can be simplified as follows:
 $$
  V(\lambda,\beta)=\frac{((\chi({\lambda})-\rho,\beta^{\vee})-m_{\beta}-2m_{2\beta})((\chi(\lambda)-\rho,\beta^{\vee})-m_{\beta}+1)}{(\beta,\beta)(\chi(\lambda)-\rho,\beta^{\vee})((\chi(\lambda)-\rho,\beta^{\vee})+1)}
  $$
  \begin{equation}
\label{SV1}
\prod_{<\alpha,\beta>>0}\frac{(\chi(\lambda)-\rho,\alpha^{\vee})-m_{\alpha}}{(\chi(\lambda)-\rho,\alpha^{\vee})+(\beta,\alpha^{\vee})-1},
  \end{equation}
 where the last  product is taken over roots $\alpha,$ which are not proportional to $\beta$.  

 \begin{thm} The super Jacobi polynomials (\ref{supjac}) satisfy the following Pieri identity 
 \begin{equation}
\label{spieri}
 2 (\sum_{i=1}^m u_i  + k^{-1}\sum_{j=1}^n v_j)\frac{\mathcal {J}_{\lambda}(u)}{\mathcal {J}_{\lambda}(0)}=\sum_{\beta\in\mathcal O}V(\lambda,\beta)\left(\frac{\mathcal {J}_{\lambda + \beta}(u)}{\mathcal{J}_{\lambda+\beta}(0)}-\frac{\mathcal {J}_{\lambda}(u)}{\mathcal {J}_{\lambda}(0)}\right),
\end{equation}
where the sum is taken over all admissible $\beta \in \mathcal O.$
\end{thm}

\begin{proof}
Let $\beta=\pm\varepsilon_i$, then a direct check shows that (\ref{SV1}) can be rewritten as
 \begin{equation}
\label{V+d}
V(\lambda, \pm\varepsilon_i)= K_{\mu}^{\pm}(\boxdot,h+n) \prod_{\Box\in\pi_{\mu}(\boxdot)} B_{\mu}^{\pm}(\Box,h+n) \prod_{\Box\in\pi'_{\lambda}(\boxdot)\setminus \pi'_{\mu}(\boxdot)} \hat B_{\lambda}^{\pm}(\Box,h),
\end{equation}
where $$h=-km-n-\frac{1}{2}p-q,$$
$\mu$ is the same as in (\ref{embedding}), $B_{\lambda}^{\pm}(\Box,h)$ and $K_{\lambda}^{\pm} (\boxdot,h)$ are given by (\ref{b+}), (\ref{b-}) and (\ref{Ki+}),(\ref{Ki-}),
$\boxdot$ denotes the added box in the $i$-th row of $\mu$, $\pi_{\mu}(\boxdot)$ is a vertical strip consisting of the boxes in the minimal rectangle containing Young diagram $\mu+\varepsilon_i$, belonging to the same column as added box but excluding the last one, $\pi'_{\lambda}(\boxdot),\,\pi'_{\mu}(\boxdot)$ are similar horizontal strips,
$\hat B_{\lambda}^{\pm}(\Box,h)= B_{\lambda'}^{\pm}(\Box,\hat h)$  is given by (\ref{b+}),(\ref{b-}), (\ref{c+}), 
(\ref{c-}), (\ref{c0}) with $\lambda$ replaced by its conjugate, $k$ by $k^{-1}$ and $\hat h$ given by (\ref{symom}). 

Similarly for $\beta=\pm\delta_j$ we have
 \begin{equation}
\label{V+delta}
V(\lambda, \pm\delta_j)= \hat K_{\tau}^{\pm}(\boxdot,h+kl(\mu)) \prod_{\Box\in\pi'_{\tau}(\boxdot)} \hat B_{\tau}^{\pm}(\Box,h+kl(\mu)) \prod_{\Box\in\pi_{\lambda}(\boxdot)\setminus \pi_{\tau}(\boxdot)} B_{\lambda}^{\pm}(\Box,h),
\end{equation}
where $\tau$ is the Young diagram $\lambda$ without the first $l(\mu)$ rows, 
$\boxdot$ is the added box in the $j$-th column of $\tau$, $\hat K_{\tau}^{\pm}(\boxdot,h))=K_{\tau'}^{\pm}(\boxdot,\hat h)$ is given by (\ref{Ki+}),(\ref{Ki-}) with $\tau$ replaced by its conjugate, $k$ by $k^{-1}$ and all other parameters given by (\ref{symom}). 

From the infinite-dimensional Pieri formula  (\ref{pieriinf}) with 
\begin{equation}
\label{V+-}
V(\lambda,h,\pm\varepsilon_i)= K_{\lambda}^{\pm}(\boxdot,h) \prod_{\Box\in\pi_{\lambda}(\boxdot)} B_{\lambda}^{\pm}(\Box,h)
\end{equation}
it follows that it is enough to prove the following 

 \begin{lemma}
 Let $\lambda$, $\mu$ and $\tau$ be as above.
We have the following identity valid for any $h$
\begin{equation}
\label{ID1}
K_{\lambda}^{\pm}(\boxdot,h)\prod_{\Box\in\pi_{\lambda}(\boxdot)\setminus \pi_{\mu}(\boxdot)} B_{\lambda}^{\pm}(\Box,h)= K_{\mu}^{\pm}(\boxdot,h+l(\tau'))\prod_{\Box\in\pi'_{\lambda}(\boxdot)\setminus \pi'_{\mu}(\boxdot)} \hat B_{\lambda}^{\pm}(\Box,h),
\end{equation}
when the box $\boxdot$ is added to $\mu.$ If $\boxdot$ is added to $\tau$ then we have
\begin{equation}
\label{ID2}
\hat K_{\tau}^{\pm}(\boxdot,h+kl(\mu))\prod_{\Box\in\pi_{\lambda}(\boxdot)\setminus \pi_{\tau}(\boxdot)} B_{\lambda}^{\pm}(\Box,h)= \hat K_{\lambda}^{\pm}(\boxdot,h)\prod_{\Box\in\pi'_{\lambda}(\boxdot)\setminus \pi'_{\tau}(\boxdot)} \hat B_{\lambda}^{\pm}(\Box,h).
\end{equation}
\end{lemma}

\noindent{\it Proof of lemma.} From Pieri formula (\ref{pieriinf}) and the symmetry of Jacobi symmetric functions (\ref{change31}) we have
$$K_{\lambda}^{\pm}(\boxdot,h) \prod_{\Box\in\pi_{\lambda}(\boxdot)} B_{\lambda}^{\pm}(\Box,h)= k^{-1}\hat K_{\lambda}^{\pm}(\boxdot,h) \prod_{\Box\in\pi'_{\lambda}(\boxdot)} \hat B_{\lambda}^{\pm}(\Box,h)
$$
and similarly
$$K_{\mu}^{\pm}(\boxdot,h+l(\tau')) \prod_{\Box\in\pi_{\mu}(\boxdot)} B_{\mu}^{\pm}(\Box,h+l(\tau'))= k^{-1}\hat K_{\mu}^{\pm}(\boxdot,h+l(\tau')) \prod_{\Box\in\pi'_{\mu}(\boxdot)} \hat B_{\mu}^{\pm}(\Box,h+l(\tau')).
$$
Dividing these two equalities and taking into account that
$$B_{\lambda}^{\pm}(\Box,h) = B_{\mu}^{\pm}(\Box, h+l(\tau'))$$ if $\Box \in \pi_{\mu}(\boxdot),$ and 
$$\hat B_{\lambda}^{\pm}(\Box,h) = \hat B_{\mu}^{\pm}(\Box, h+l(\tau'))$$ if $\Box \in \pi'_{\mu}(\boxdot)$  
we get (\ref{ID1}). The second identity can be proved similarly. This proves the lemma and the theorem.  
\end{proof}

One can interpret the formula as the  {\it bispectrality} between the deformed $BC(m,n)$ CMS operator (\ref{bcnm}) and the following {\it deformed rational Koornwinder operator}
 \begin{equation}
\label{defKoorn}
\mathcal D^{(m,n)}= \sum_{\beta\in\mathcal O}W(z,\beta)(T_{\beta}-1),
 \end{equation}
where $$z=\sum_{i=1}^m x_i \varepsilon_i +\sum_{i=1}^n y_j \delta_j,$$ 
$T_{\beta}: f(z) \rightarrow f(z+\beta)$ is the shift operator and 
\begin{equation}
\label{SW1}
W(z,\beta)=\frac{((z,\beta^{\vee})-m_{\beta}-2m_{2\beta})((z,\beta^{\vee})-m_{\beta}+1)}{(\beta,\beta)(z,\beta^{\vee})((z,\beta^{\vee})+1)}
  \prod_{<\alpha,\beta>>0}\frac{(z,\alpha^{\vee})-m_{\alpha}}{(z,\alpha^{\vee})+(\beta,\alpha^{\vee})-1},
 \end{equation}
 where again the  product is taken over roots $\alpha,$ which are not proportional to $\beta$.  
 More explicitly, 
 \begin{equation}
\label{defKoornex}
 \mathcal D^{(m,n)}= \sum_{i=1}^m [W^+_{i,x}(T_{i,x}-1) + W^-_{i,x}(T_{i,x}^{-1}-1)] + \sum_{j=1}^n [W^+_{j,y}(T_{j,y}-1) + W^-_{j,y}(T_{j,y}^{-1}-1)],
 \end{equation}
where $T_{i,x}, T_{j,y}$ are the shifts by 1 in $i$-th $x$-direction and $j$-th $y$-direction respectively,
$$
W^+_{i,x}=(1-\frac{p+2q}{2x_i})(1-\frac{p}{2x_i+1})\prod_{j\neq i}^m(1-\frac{k}{x_i-x_j})(1-\frac{k}{x_i+x_j})$$
$$
\prod_{j=1}^n \left(1-\frac{1}{x_i+ky_j+\frac{1-k}{2}}\right)\left(1-\frac{1}{x_i-ky_j+\frac{1-k}{2}}\right),
$$
$$
W^-_{i,x}=(1+\frac{p+2q}{2x_i})(1+\frac{p}{2x_i-1})\prod_{j\neq i}^m(1+\frac{k}{x_i-x_j})(1+\frac{k}{x_i+x_j})$$
$$
\prod_{j=1}^n \left(1+\frac{1}{x_i+ky_j-\frac{1-k}{2}}\right)\left(1+\frac{1}{x_i-ky_j-\frac{1-k}{2}}\right),
$$
$$
W^+_{j,y}=\frac{1}{k}(1-\frac{r+2s}{2y_j})(1-\frac{r}{2y_j+1})\prod_{i\neq j}^n(1-\frac{k^{-1}}{y_j-y_i})(1-\frac{k^{-1}}{y_j+y_i})
$$
$$
\prod_{i=1}^m \left(1-\frac{1}{y_j+k^{-1}x_i+\frac{1-k^{-1}}{2}}\right)\left(1-\frac{1}{y_j-k^{-1}x_i+\frac{1-k^{-1}}{2}}\right),
$$
$$
W^-_{j,y}=\frac{1}{k} (1+\frac{r+2s}{2y_j})(1+\frac{r}{2y_j-1})\prod_{i\neq j}^n(1+\frac{k^{-1}}{y_j-y_i})(1+\frac{k^{-1}}{y_j+y_i})$$
$$
\prod_{i=1}^m \left(1+\frac{1}{y_j+k^{-1}x_i-\frac{1-k^{-1}}{2}}\right) \left(1+\frac{1}{y_j-k^{-1}x_i-\frac{1-k^{-1}}{2}}\right).
$$
As we have mentioned before in the special cases $n=0$ and $n=1$ the bispectrality was established earlier in \cite{Cha} and \cite{F2}.

We would like to emphasize that all the formulas for super Jacobi polynomials can essentially be found from the usual case simply replacing the $BC_N$ root system by its $BC(m,n)$ deformed version. This fact is another evidence of the importance of the notion of the deformed root systems.

\section{Acknowledgements}

We are grateful to Felipe van Diejen and Andrei Okounkov for useful discussions and comments.

This work has been partially supported by EPSRC (grant EP/E004008/1) and by the
European Union through the FP6 Marie Curie RTN ENIGMA (contract
number MRTN-CT-2004-5652) and through ESF programme MISGAM.

\end{document}